\documentclass{nature_plusfigure}
\usepackage{graphicx}
\usepackage{subcaption}
\usepackage{aas_macros,amsmath,amssymb,xcolor,url}
\usepackage{multirow}
\usepackage{longtable,hhline}
\usepackage{upgreek}
\usepackage{url}
\usepackage{hyperref}

\usepackage[none]{hyphenat}
\usepackage{multibib}
\newcites{New}{References}

\usepackage{xspace}
\newcommand\eyj{SN~2020eyj\xspace}
\newcommand{\msun}{$M_{\odot}$}

\usepackage[symbol]{footmisc}

\usepackage{cleveref}
\usepackage{subcaption} 

\title{A radio-detected Type Ia supernova with helium-rich circumstellar material}

\author{Erik C. Kool$^{1}$, Joel Johansson$^{2}$, Jesper Sollerman$^{1}$, Javier Mold\'on$^{3,4}$, Takashi J. Moriya$^{5,6}$, Seppo Mattila$^{7,8}$, Steve Schulze$^{2}$, Laura Chomiuk$^{9}$, Miguel P\'erez-Torres$^{3,10}$, Chelsea Harris$^{9}$, Peter Lundqvist$^{1}$, Matthew Graham$^{11}$, Sheng Yang$^{1}$, Daniel A. Perley$^{12}$, Nora Linn Strotjohann$^{13}$, Christoffer Fremling$^{11}$, Avishay Gal-Yam$^{13}$, Jeremy Lezmy$^{14}$, Kate Maguire$^{15}$, Conor Omand$^{1}$, Mathew Smith$^{14,16}$, Igor Andreoni$^{17,18,19}$, Eric C. Bellm$^{20}$, Joshua~S. Bloom$^{21,22}$, Kishalay De$^{23}$, Steven L. Groom$^{24}$, Mansi M. Kasliwal$^{11}$, Frank J. Masci$^{24}$, Michael S. Medford$^{21,22}$, Sungmin Park$^{25}$, Josiah Purdum$^{26}$, Thomas M. Reynolds$^{27}$, Reed Riddle$^{11}$, Estelle Robert$^{14}$, Stuart D. Ryder$^{28,29}$, Yashvi Sharma$^{11}$, Daniel Stern$^{30}$}
	
\begin{document}

\maketitle

\begin{small}
\begin{affiliations}
\label{sec:affiliations}

\item The Oskar Klein Centre, Department of Astronomy, Stockholm University, AlbaNova, SE-10691, Stockholm, Sweden
\item The Oskar Klein Centre, Department of Physics, Stockholm University, AlbaNova, SE-10691, Stockholm, Sweden
\item Instituto de Astrof\'isica de Andaluc\'ia, Consejo Superior de Investigaciones Cient\'ificas (CSIC), Glorieta de la Astronom\'ia s/n, 18008 Granada, Spain
\item Jodrell Bank Centre for Astrophysics, School of Physics and Astronomy, The University of Manchester, Manchester M13 9PL, UK
\item National Astronomical Observatory of Japan, National Institutes of Natural Sciences, 2-21-1 Osawa, Mitaka, Tokyo 181-8588, Japan
\item School of Physics and Astronomy, Faculty of Science, Monash University, Clayton, Victoria 3800, Australia
\item Tuorla Observatory, Department of Physics and Astronomy, University of Turku, FI-20014 Turku, Finland
\item School of Sciences, European University Cyprus, Diogenes street, Engomi, 1516 Nicosia, Cyprus
\item Center for Data Intensive and Time Domain Astronomy, Department of Physics and Astronomy, Michigan State University, East Lansing, MI 48824, USA
\item Facultad de Ciencias, Universidad de Zaragoza, Pedro Cerbuna 12, E-50009 Zaragoza, Spain
\item Division of Physics, Mathematics and Astronomy, California Institute of Technology, Pasadena, CA, USA
\item Astrophysics Research Institute, Liverpool John Moores University, IC2, Liverpool Science Park, 146 Browlow Hill, Liverpool L3 5RF, UK
\item Department of Particle Physics and Astrophysics, Weizmann Institute of Science, 76100 Rehovot, Israel
\item Univ Lyon, Univ Claude Bernard Lyon 1, CNRS/IN2P3, IP2I Lyon, UMR 5822, F-69622, Villeurbanne, France
\item School of Physics, Trinity College Dublin, The University of Dublin, Dublin, Ireland
\item School of Physics and Astronomy, University of Southampton, Southampton, Hampshire SO17 1BJ, UK
\item Joint Space-Science Institute, University of Maryland, College Park, MD 20742, USA
\item Department of Astronomy, University of Maryland, College Park, MD 20742, USA
\item Astrophysics Science Division, NASA Goddard Space Flight Center, Mail Code 661, Greenbelt, MD 20771, USA
\item DIRAC Institute, Department of Astronomy, University of Washington, 3910 15th Avenue NE, Seattle, WA 98195, USA
\item Department of Astronomy, University of California, Berkeley, CA 94720-3411, USA
\item Lawrence Berkeley National Laboratory, 1 Cyclotron Road, MS 50B-4206, Berkeley, CA 94720, USA
\item Kavli Institute for Astrophysics and Space Research, Massachusetts Institute of Technology, Cambridge, MA 02139, USA
\item IPAC, California Institute of Technology, 1200 E. California Blvd, Pasadena, CA 91125, USA
\item Ulsan National Institute of Science and Technology, 50 UNIST-gil, Ulsan, South Korea
\item Caltech Optical Observatories, California Institute of Technology, Pasadena, CA 91125, USA
\item The Cosmic Dawn Center (DAWN), Niels Bohr Institute, University of Copenhagen, R\aa dmandsgade 62, DK-2200 K\o benhavn N, Denmark
\item School of Mathematical and Physical Sciences, Macquarie University, NSW 2109, Australia
\item Astronomy, Astrophysics and Astrophotonics Research Centre, Macquarie University, Sydney, NSW 2109, Australia
\item Jet Propulsion Laboratory, California Institute of Technology, 4800 Oak Grove Drive, Pasadena, CA 91109, USA

\end{affiliations}
\end{small}

\begin{abstract}

Type Ia supernovae (SNe Ia) are thermonuclear explosions of degenerate white dwarf (WD) stars destabilized by mass accretion from a companion star \cite{Whelan1973}, but the nature of their progenitors remains poorly understood. A way to discriminate between progenitor systems is through radio observations; a non-degenerate companion star is expected to lose material through winds \cite{Seaquist1990} or binary interaction \cite{Hachisu1996} prior to explosion, and the SN ejecta crashing into this nearby circumstellar material (CSM) should result in radio synchrotron emission. However, despite extensive efforts, no SN Ia has ever been detected at radio wavelengths, which suggests a clean environment and a companion star that is itself a degenerate WD star \cite{Horesh2012,pereztorres2014}. Here we report on the study of \eyj, a SN Ia showing helium-rich CSM, as revealed by its spectral features, infrared emission and, for the first time in a SN Ia, a radio counterpart. Based on our modeling, we conclude the CSM likely originates from a single-degenerate (SD) binary system where a WD accretes material from a helium donor star, an often hypothesized formation channel for SNe Ia \cite{wang2009a,Ruiter2009}. We describe how comprehensive radio follow-up of \eyj-like SNe Ia can improve the constraints on their progenitor systems. 


\end{abstract}

\begin{figure}[t!] 
  \centering
  \includegraphics[width=0.75\columnwidth]{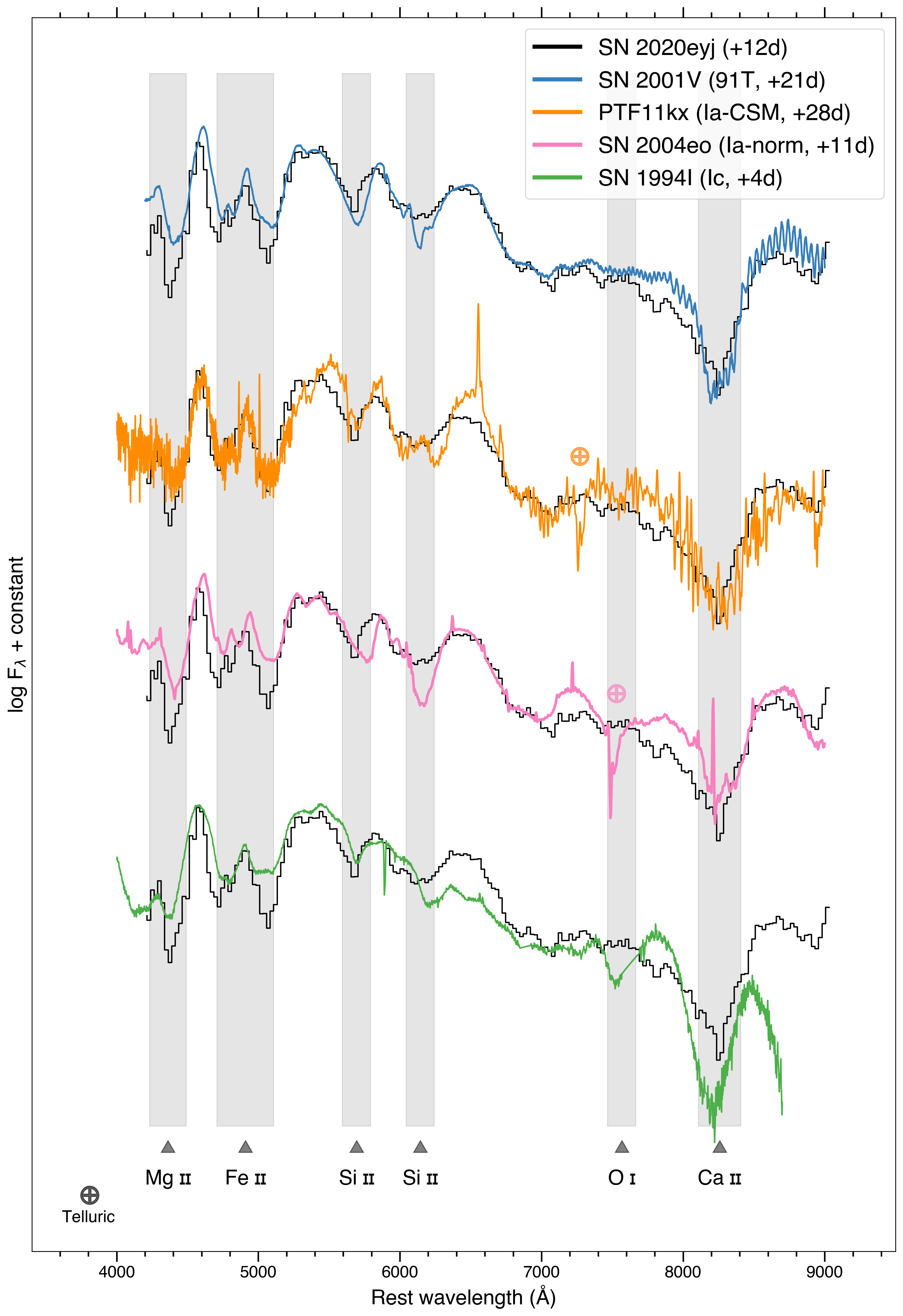}
  \caption{{\bf The first spectrum of \eyj is consistent with a Type Ia(-CSM)}. The SEDM classification spectrum of \eyj, obtained $\sim$12 days after peak and shown in black, is compared with Type Ia-91T SN 2001V, Type Ia-CSM PTF11kx, Type Ia SN 2004eo and Type Ic SN 1994I. Phases are relative to peak, which in the case of \eyj has an uncertainty of a couple of days. A number of important absorption features are indicated at the expected wavelengths. Notably, the spectrum of \eyj lacks any sign of O\,{\sc i} 7774 \AA\ absorption. Spectra have been corrected for MW reddening.}
  \label{fig:classificationspectrum}
\end{figure}

\eyj was first detected on 2020 March 07 UT (MJD = 58915.12, Sect.~\ref{sec:discovery}), at $\alpha=11^{h}11^{m}47.19^{s}$, $\delta=$ +29\textdegree23$^{\prime}$06.5$^{\prime\prime}$ (J2000). The SN was classified as a SN Ia \cite{Dahiwale2020} based on a low-resolution spectrum obtained on 2020 April 2, 25 days after the first detection. Comparisons with Type Ia and Ibc spectra from the literature support the SN Ia classification (Sect.~\ref{sec:snia} and Fig.~\ref{fig:classificationspectrum}). Unusual evolution of the later light curve prompted us to obtain a second spectrum on 2020 July 20, 131 days after first detection. The second spectrum was very similar to those of Type Ibn SNe (SNe Ibn), which are SNe that interact with helium-rich CSM and have spectra characterized by narrow ($\sim$few$\times 10^{3}$ km s$^{-1}$) He\,{\sc i} emission lines while showing little to no H\,{\sc i} \cite{foley2007,pastorello2007b}. 

\begin{figure}[t!] 
  \centering
  \includegraphics[width=0.99\columnwidth]{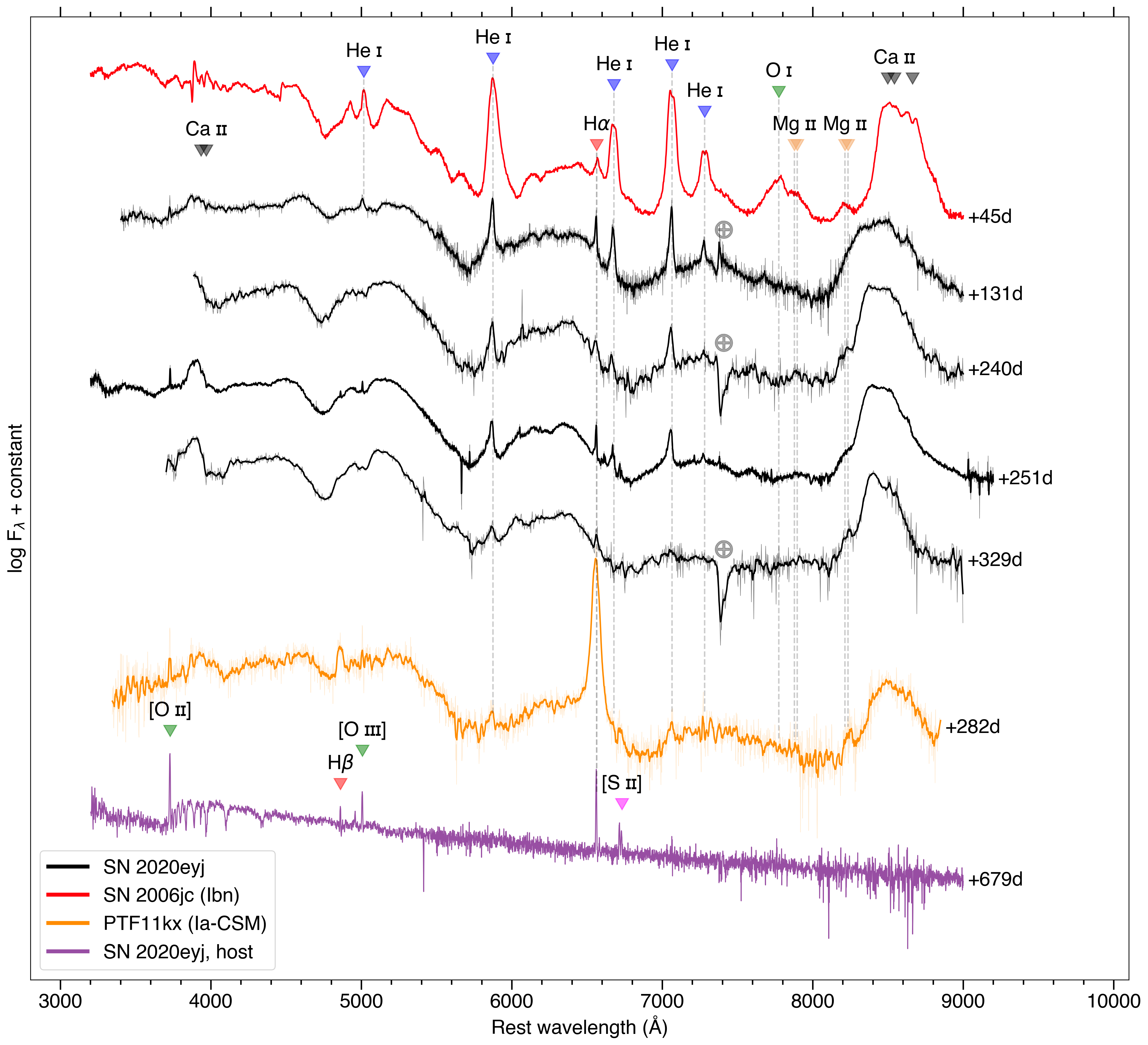}
  \caption{{\bf The spectra of \eyj in the tail phase are dominated by CSM interaction}. The spectra of \eyj at late phases (in black) are compared with the prototypical Type Ibn SN 2006jc and the Type Ia-CSM SN PTF11kx. The spectra show features common to SNe Ia-CSM, such as the quasi-continuum blueward of 5700~\AA\ and broad Ca\,{\sc ii} emission. The main SN emission features are identified in the top spectrum. The emission lines in \eyj show strong asymmetry, with attenuated red wings (Fig.~\ref{fig:emlines}). The bottom spectrum is of the host of \eyj, obtained at 678 days, some 300 days after the SN had faded below the detection limit of ZTF. Some unresolved galaxy lines are marked. Phases are relative to first detection, which in the case of SN 2006jc was at or after the peak. Spectra have been corrected for MW reddening.}
  \label{fig:laterspectra}
\end{figure}

Based on the late time (tail phase) CSM-interaction dominated spectra (Fig.~\ref{fig:laterspectra}), \eyj falls in the category of the rare subclass of SNe Ia that show evidence of CSM interaction in their optical spectra (SNe Ia-CSM; \cite{Silverman2013}). The narrow emission lines in the spectra of such interacting SNe arise from shock interaction between the fast-moving SN ejecta and the slow-moving CSM \cite{smith2017}. SNe Ia-CSM are strong contenders for the single degenerate (SD) SN Ia formation channel on account of the CSM, which is commonly assumed to originate from a non-degenerate donor star through stellar or accretion winds. Prior to \eyj, all the discovered SNe Ia-CSM exhibited prominent Balmer emission lines and only weak He emission features \cite{Silverman2013}. 

\begin{figure}[t!] 
  \centering
  \includegraphics[width=0.99\columnwidth]{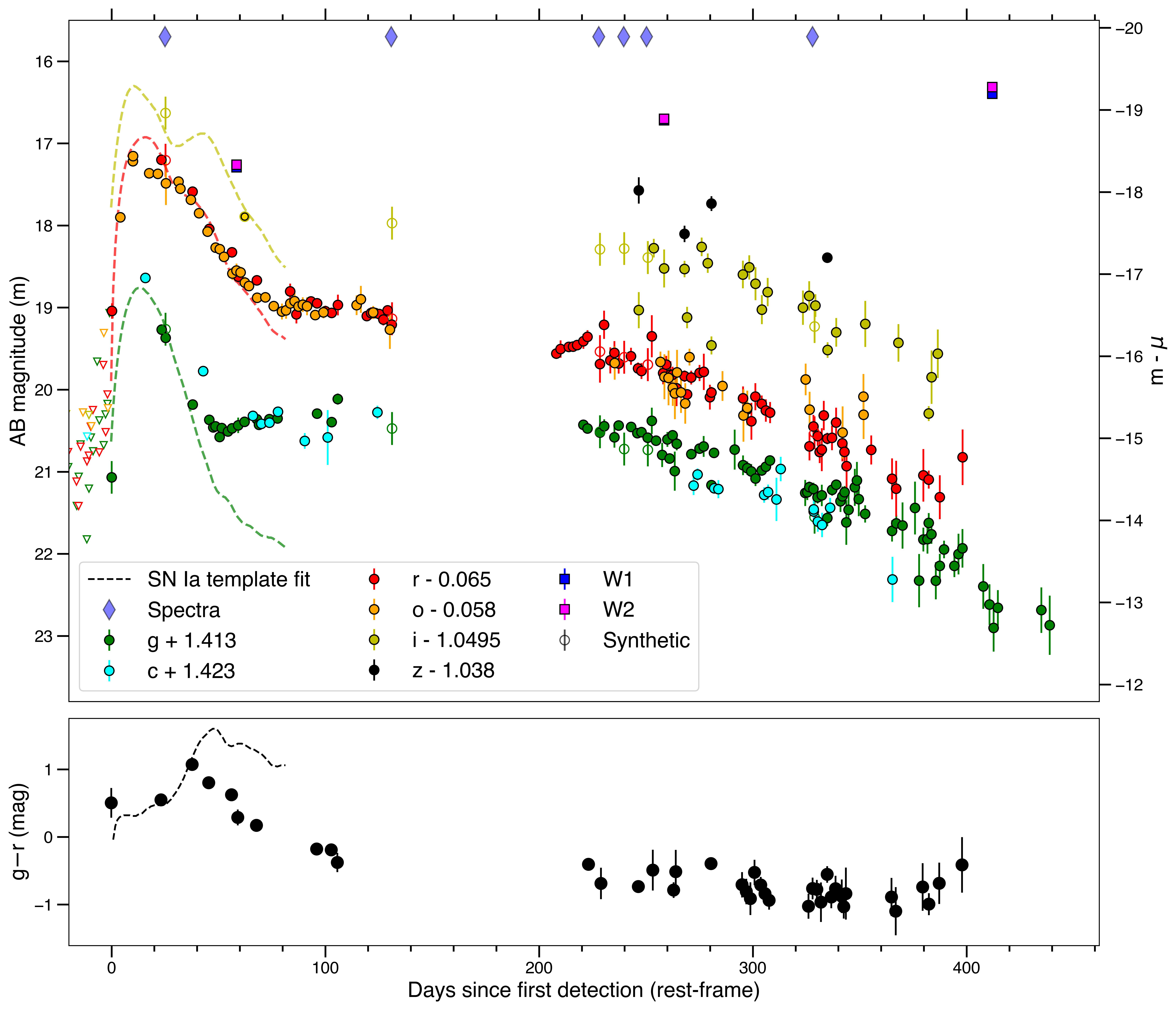}
  \caption{{\bf The multi-band light curve of \eyj can be divided into a diffusion peak phase and a long-lived interaction-powered tail phase}. The light curves of \eyj are shown with over-plotted SN Ia template fits to the initial peak (Sect.~\ref{sec:lightcurve}). The most recent mid-IR epoch (W1 and W2) is outside the date range plotted here, and is shown in Fig.~\ref{fig:wise} instead. Open circles indicate synthetic photometry derived from the spectra. Phase is in rest-frame days since first detection. Apparent magnitudes on the left y-axis, absolute magnitudes on the right y-axis, where $\mu$ is the distance modulus. Non-detections with 5$\sigma$ upper limits are indicated by triangles. The photometry has been binned into one-night bins and has been corrected for MW reddening. The diamond markers on top indicate the epochs of spectroscopy. The bottom panel shows the $g-r$ color for the nights in which both $g$ and $r$ photometry was obtained, with overplotted the color evolution of a typical SN Ia. The error bars represent 1$\sigma$ uncertainties.}
  \label{fig:lightcurve}
\end{figure}

Typically, CSM interaction contributes significantly to or even dominates the spectral and light curve evolution of SNe Ia-CSM from the start, hindering unambiguous classification as SNe Ia \cite{Leloudas2015}. However, in some rare cases SNe Ia-CSM have shown a delay in CSM interaction \cite{hamuy2003,dilday2012,graham2019a}, suggesting the CSM was located far ($>10^{15}$ cm) from the binary system at the time of explosion. Notably, PTF11kx cemented SNe Ia-CSM as a bona fide SN Ia subclass by virtue of a delay of $\sim$60 days, allowing for an indisputable SN Ia classification prior to CSM interaction \cite{dilday2012}. \eyj follows a similar evolution as PTF11kx, initially showing a typical SN Ia bell-shaped light curve (Fig.~\ref{fig:lightcurve}) and a spectrum consistent with a SN Ia of the 91T subgroup \cite{Branch2006} without clear evidence for CSM interaction (Fig.~\ref{fig:classificationspectrum}). Then, at 50 days after first detection, the $g$-band light curve of \eyj diverges from a steady decline into a plateau that lasts $\sim$200 days. Such an evolution and color change is not expected for a normal SN Ia (Fig.~\ref{fig:lightcurve}), but is driven by the emergence of spectral features associated with CSM interaction (Sect.~\ref{sec:lightcurve}). We interpret the start of the plateau at 50 days as the epoch when CSM interaction starts to contribute significantly or to even dominate the light curve of \eyj. Assuming a SN ejecta velocity of $10^4$ km s$^{-1}$ \cite{Wilk2018}, the delay corresponds to an inner boundary to the CSM of $\sim4\times10^{15}$ cm. Save for the presence of He emission lines, the late-time spectra of \eyj are typical for the SN Ia-CSM class, with prominent broad Ca\,{\sc ii} emission from the near-infrared (near-IR) triplet and without any sign of O\,{\sc i} $\lambda$7774 emission (Fig.~\ref{fig:laterspectra}). The compact and star-forming host galaxy of \eyj (Sect.~\ref{sec:host}) is also consistent with those of other SNe Ia-CSM \cite{Silverman2013}.

Despite the similarities between \eyj and other SNe Ia-CSM, the presence of He\,{\sc i} lines and absence of prominent H\,{\sc i} lines remains a striking difference with profound implications for the progenitor system. As H\,{\sc i} is easier to ionize than He\,{\sc i}, the absence of the lines indicates that the CSM around \eyj, and thus the companion star, is He-rich and H-poor. While the late-time spectra of \eyj are similar to those of SNe Ibn, these SNe are presumed to arise from the core collapse of massive ($>$ 10 \msun) stars \cite{foley2007,pastorello2008I,Dessart2022}, which are unlikely to be in a binary system with a WD, as they would undergo core collapse long before the WD formed.
A merger involving a degenerate He WD donor star is also disfavored, because in such merger models only a small amount of unburned He ($\sim0.03$ \msun \cite{Boyle2017}) is present close to ($\lesssim10^{12}$ cm) the WD \cite{Kromer2010}, whereas the CSM around \eyj resides at $>10^{15}$ cm. Instead, a strong candidate for the donor star in the \eyj progenitor system is a non-degenerate He star (initial mass 1$-$2 \msun, e.g. \cite{Yoon2003}). WD + He star systems can be formed via binary evolution \cite{Iben1994}, and this SD channel for SNe Ia has garnered recent interest because the very restrictive limits placed by radio non-detections and deep optical imaging \cite{Li2011} that exclude most H-rich donor star models, still allow for low CSM density WD + He star systems \cite{Li2011,Moriya2019}. The possible detection in pre-explosion HST imaging of the progenitor system of the Type Iax (SNe Ia similar to SN~2002cx \cite{Li2003}) SN 2012Z, a blue compact source interpreted as a He-star donor \cite{McCully2014}, has further strengthened this hypothesis, although the thermonuclear nature of Type Iax SNe is debated \cite{Valenti2009}.

\begin{figure}[t!]
  \centering
  \includegraphics[width=0.9\columnwidth]{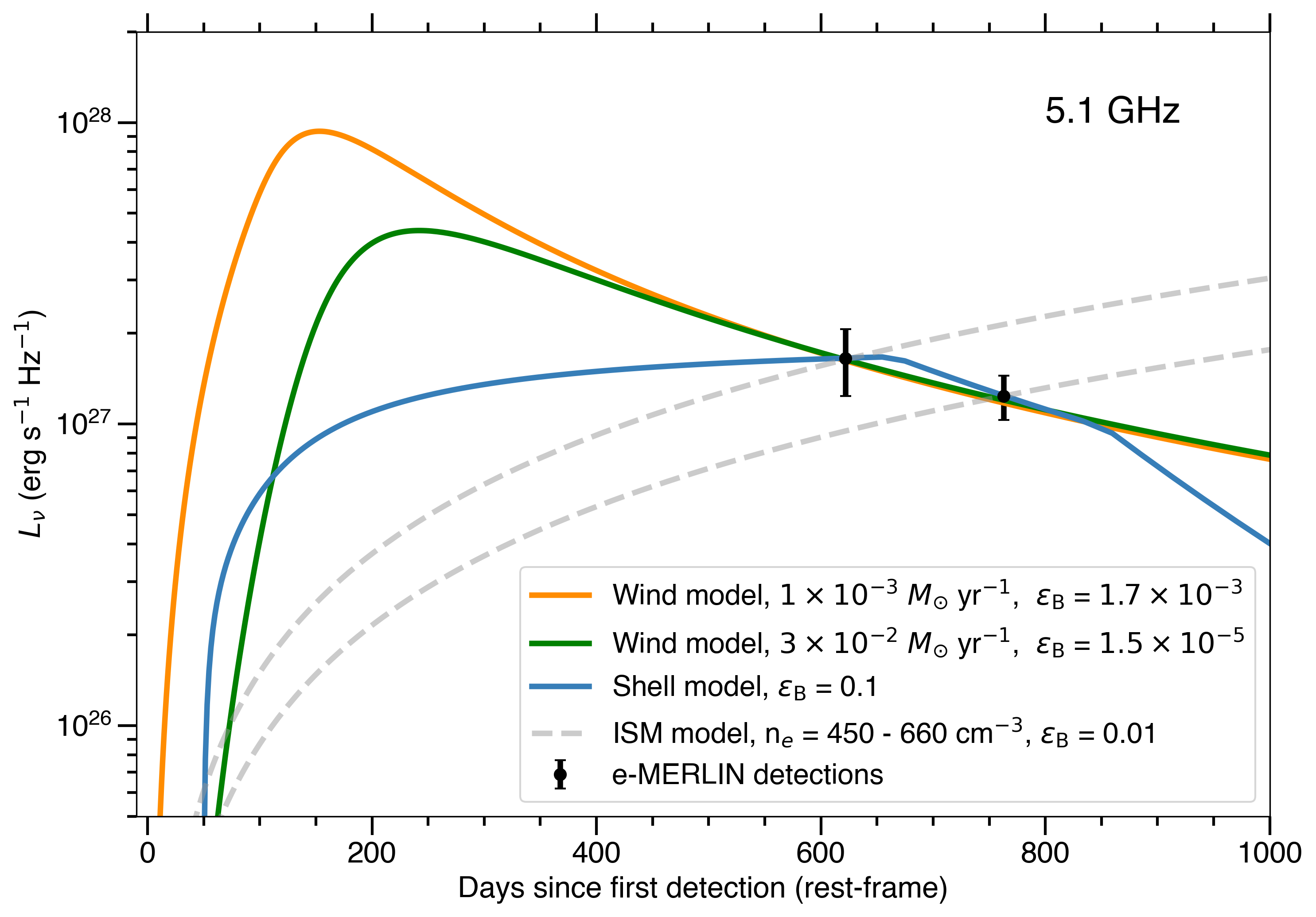} 
  \caption{{\bf The radio detections of \eyj at 5.1 GHz can be reconciled with CSM interaction
  .} For the wind model, where the CSM follows a density profile of $\rho \propto r^{-2}$, we assume a pre-SN wind velocity of 1000 km s$^{-1}$ and adopt a mass-transfer rate as inferred from fitting the bolometric light curve of \eyj. Depending on the level of line of sight extinction affecting the bolometric light curve (Sect.~\ref{sec:wind}), the wind model fits the observations (in black, with 1$\sigma$ uncertainties) well for the microphysics parameter $\epsilon_{\textrm{B}} = 1.7\times10^{-3}$ ($1.5\times10^{-5}$), and a CSM mass of $M_{\mathrm{csm}} = 0.3$ \msun\ (1~\msun) within $10^{17}$ cm (Sect.~\ref{sec:wind}) when E(B-V) = 0 mag (0.5 mag). For the shell model, where the CSM is concentrated in a constant density CSM shell, we assume $\epsilon_{\textrm{B}}$ = 0.1, and obtain a best estimate for the CSM mass of $M_{\mathrm{csm}} = 0.36$ \msun\ and a CSM interaction end time of $t_{\mathrm{end}} = 665$ days (a width of $8.6\times10^{16}~\mathrm{cm}$; Sect.~\ref{sec:shells}). In both the wind and shell model fits, $\epsilon_{\textrm{e}}$ = 0.1 is assumed. We also show radio light curves from a model involving a DD Type Ia SN interacting with the ISM (Sect.~\ref{sec:dd_radio}). In order to fit the individual radio detections, this model requires unusually high ISM densities, and neither fit reproduces the observed decline in flux, ruling out the DD scenario.}
  \label{fig:radio_comparison}
\end{figure}

The CSM interaction in \eyj is also confirmed, for the first time in a Type Ia SN, through the detection of a radio counterpart, at a frequency of 5.1 GHz at $605$ and $741$ days after the first detection (Sect.~\ref{sec:radio}). Follow-up in the X-rays did not yield a detection (Sect.~\ref{sec:xray}). We model the radio synchrotron emission, which results from the shock interaction between the ejecta and the CSM, assuming two basic CSM configurations expected in a SD progenitor system; a constant density shell, and a wind-like density profile with density $\rho \propto r^{-2}$ (Fig.~\ref{fig:radio_comparison}). A constant density shell could result from a mass ejection event such as a nova, whereas a wind-like CSM profile would be expected from an optically thick wind, where the mass-transfer rate from the donor star to the WD exceeds the maximum accretion rate of He-rich material that the WD can burn on its surface \cite{Nomoto1982b,Moriya2019}. In addition to CSM material arising from a SD scenario, we consider synchrotron emission resulting from the interaction of a SN Ia from a double degenerate (DD) white dwarf merger interacting with the local interstellar medium (ISM) \cite{Kundu2017}. For the SD shell model, the radio detections are best explained with a CSM mass of $M_{\mathrm{csm}} = 0.36$ \msun\ (Sect.~\ref{sec:shells}), with the expectation that the radio light curve will start to drop off quite rapidly at $\sim900$ days. For the SD optically thick wind model, the bolometric light curve tail (Sect.~\ref{sec:bolometric}) and radio detections of \eyj are well fitted with a mass-transfer rate of $10^{-3} - 10^{-2}$~\msun~yr$^{-1}$, microphysics parameter $\epsilon_{\textrm{B}}$ = $10^{-5} - 10^{-3}$, and a CSM mass within $10^{17}$ cm of $M_{\mathrm{csm}} = 0.3 - 1.0$ \msun. The DD ISM model (the striped lines in Fig.~\ref{fig:radio_comparison}) requires unusually high ISM densities and does not recover the observed decline in flux, ruling out the DD formation channel for \eyj (Sect.~\ref{sec:dd_radio}). The best fit radio light curves of the shell and wind models differ in particular at early phases (Fig.~\ref{fig:radio_comparison}), but no radio data were obtained at these epochs. Instead, multi-frequency monitoring of the radio counterpart of \eyj until late phases ($>1000$ days) will allow to discriminate between the rapid drop-off of the shell model, and a shallower decline expected in the case of a wind-like CSM.

A viable progenitor scenario for \eyj needs to explain not only the presence and properties of a He-rich CSM, but also its detached configuration. For the delayed Type Ia-CSM SN~2002ic, the CSM free cavity was attributed to a possible drop-off in mass-transfer rate or the emergence of a low-density fast wind evacuating the CSM \cite{woodvasey2004}. In the case of PTF11kx, the delayed CSM interaction was explained by a scenario involving a symbiotic nova progenitor, where recurrent novae on the surface of the WD sweep up the wind-deposited CSM into shells \cite{dilday2012}. \eyj shows strong similarities to PTF11kx, which may hint at a common progenitor scenario. Their light curves are virtually identical up until day 50 (Fig.~\ref{fig:lightcurve_fits}) with rise times of $\sim$14 days in $g$ band, which is fast for a SN Ia \cite{Miller2020}. And, except for the nature of the narrow emission lines, both SNe have similar spectra throughout their evolution (Figs.~\ref{fig:classificationspectrum} and \ref{fig:laterspectra}). For \eyj, a nova progenitor could look like V445 Puppis (V445 Pup, Sect.~\ref{sec:v445_pup}), the only known nova system that showed He-rich, but H-free, ejecta \cite{Ashok2003,Woudt2009}. Notably, the V445 Pup system is considered a prime candidate progenitor system for the He star + WD SN Ia channel, as it is claimed to be host to a WD with a mass close to the Chandrasekhar limit \cite{Kato2008}. Additionally, a prominent carbon rich equatorial dusty disc like the one in V445 Pup \cite{Ashok2003,Woudt2009} could explain (Sect.~\ref{sec:v445_pup}) the luminous IR counterpart of \eyj (Fig.~\ref{fig:wise}), which we attribute to an IR echo from radiatively heated pre-existing dust with a dust mass of order $10^{-2}$~\msun~ (Sect.~\ref{sec:infrared}). The initial models invoking recurrent novae for the origin of PTF11kx \cite{dilday2012} were challenged by the CSM masses involved \cite{Soker2013}, which were too large by orders of magnitude for symbiotic nova mass build-up models \cite{Moore2012}. Similarly, the mass resulting from a V445 Pup-like nova outburst ($\lesssim10^{-3}$\msun; Sect.~\ref{sec:lightcurve}) is insufficient to explain the CSM mass observed in \eyj. However, a recent study of the radio evolution of V445 Pup suggests that the equatorial disk could have pre-dated (and survived) the nova outburst \cite{Nyamai2021}, which would allow for mass build-up in the disc between nova eruptions. This scenario would require the SN to occur soon after the nova outburst, and before the resumption of mass-transfer between the donor and WD reforms the disc at small radii. We note that a nova similar to the year 2000 event of V445 Pup would not have been detectable at the distance of \eyj (Sect.~\ref{sec:precursor}). 

\eyj represents the first observational example of the previously speculated class of SNe Ia-He CSM \cite{Nomoto2018}. The presence of a dense CSM, supported by a radio detection, offers strong evidence for the SD scenario for \eyj, in particular for the He star + WD formation channel. It is estimated $\sim$10\% of all SD Type Ia SNe arise from this channel \cite{Ruiter2009}, which is likely the dominant source of SNe Ia with short delay times \cite{wang2009b}. Understanding the timescale of SN Ia activity is important for the chemical evolution of galaxies. The confirmed presence of a He-rich CSM in a SN Ia system also impacts SN Ia explosion modeling, as He plays a vital role in double detonation models where the WD explosion is triggered by the ignition of a massive ($\lesssim$ 0.2 \msun) He shell on its surface \cite{Nomoto1982b}.
Constraining the rate of SNe Ia similar to \eyj would require systematic spectroscopic follow up of SNe Ia with long-lived light curves, as currently monitoring often stops after a seemingly normal SN Ia has been classified. Observational properties which \eyj shares with its H-analog PTF11kx, such as a fast rise and a 91T-like peak spectrum, can potentially guide such follow up efforts and allow for the discovery and study of more \eyj-like SNe Ia, including at radio wavelengths.

\clearpage

\clearpage
\clearpage

\clearpage

\clearpage

\begin{extended_data}

\renewcommand{\thefigure}{\arabic{figure}~Extended~Data}
\renewcommand{\thefigure}{Extended Data Figure \arabic{figure}}
\renewcommand{\figurename}{}
\setcounter{figure}{0}

\renewcommand{\thetable}{\arabic{table}~Extended~Data}
\renewcommand{\thetable}{Extended Data Table \arabic{table}}
\renewcommand{\tablename}{}
\setcounter{table}{0}

\begin{table}
\caption{\label{tab:properties} Key characteristics of \eyj}
\centering
\begin{tabular}{l|c} \hline
$\alpha$ (J2000)                        &   $11^{h}11^{m}47.19^{s}$ \\
$\delta$  (J2000)                       &   +29\textdegree23$^{\prime}$06.5$^{\prime\prime}$ \\
Luminosity distance (Mpc)               &   131.4\\
First detection epoch (MJD)             &   58915.212\\
Peak epoch (fit, MJD)                   &   58929 $\pm$ 2\\
Redshift                                &   $0.0297 \pm 0.0001$\\
E(B$-$V$)_{\textrm{MW}}$ (mag)          &  0.024\\
\hline
\end{tabular}
\end{table}

\clearpage

\begin{figure*}
\centering
    \begin{subfigure}[t]{0.32\textwidth}
         \centering
         \includegraphics[width=0.99\textwidth]{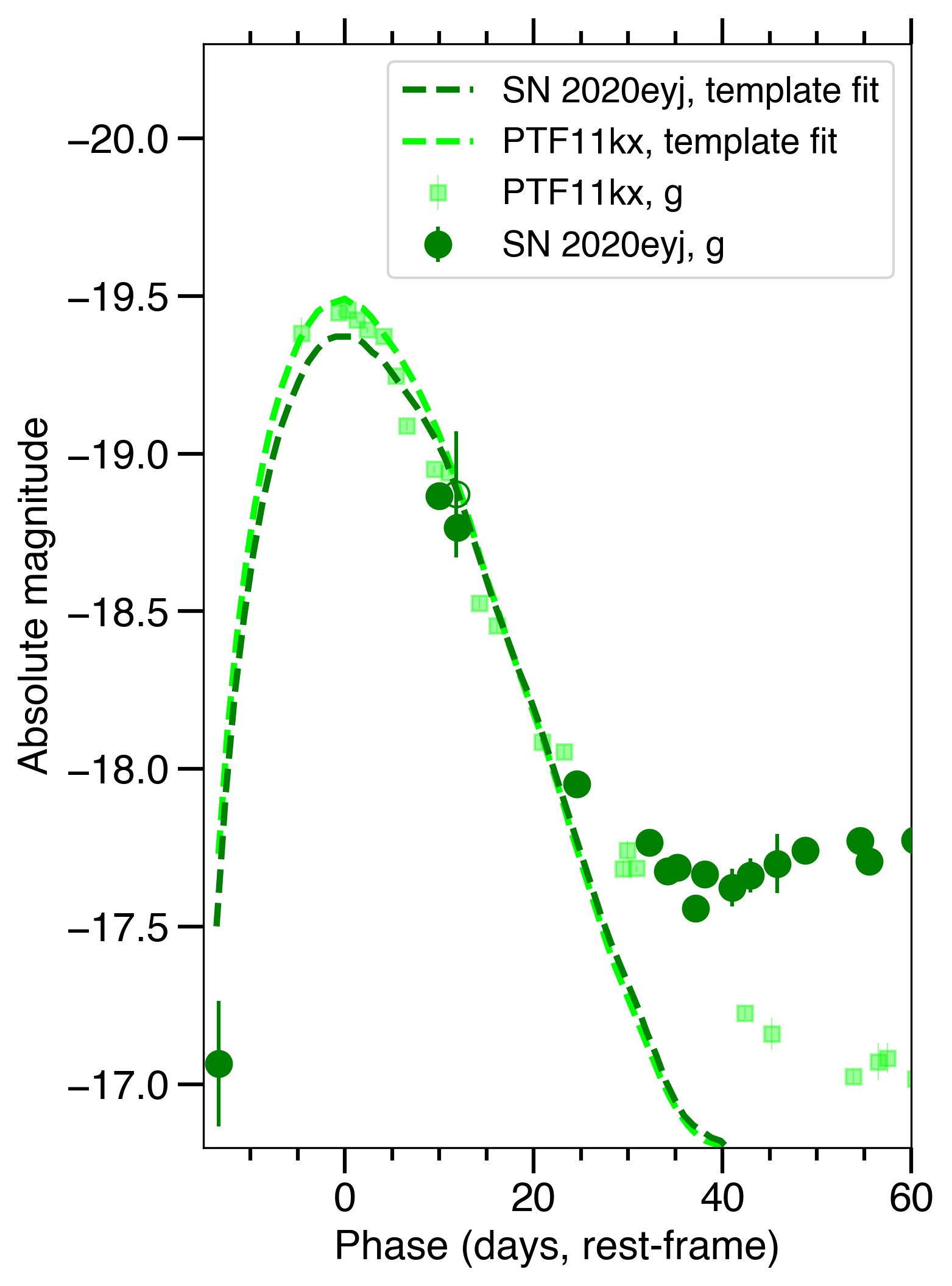}
         \caption{$g$ band}
     \end{subfigure}
    \begin{subfigure}[t]{0.32\textwidth}
         \centering
         \includegraphics[width=0.99\textwidth]{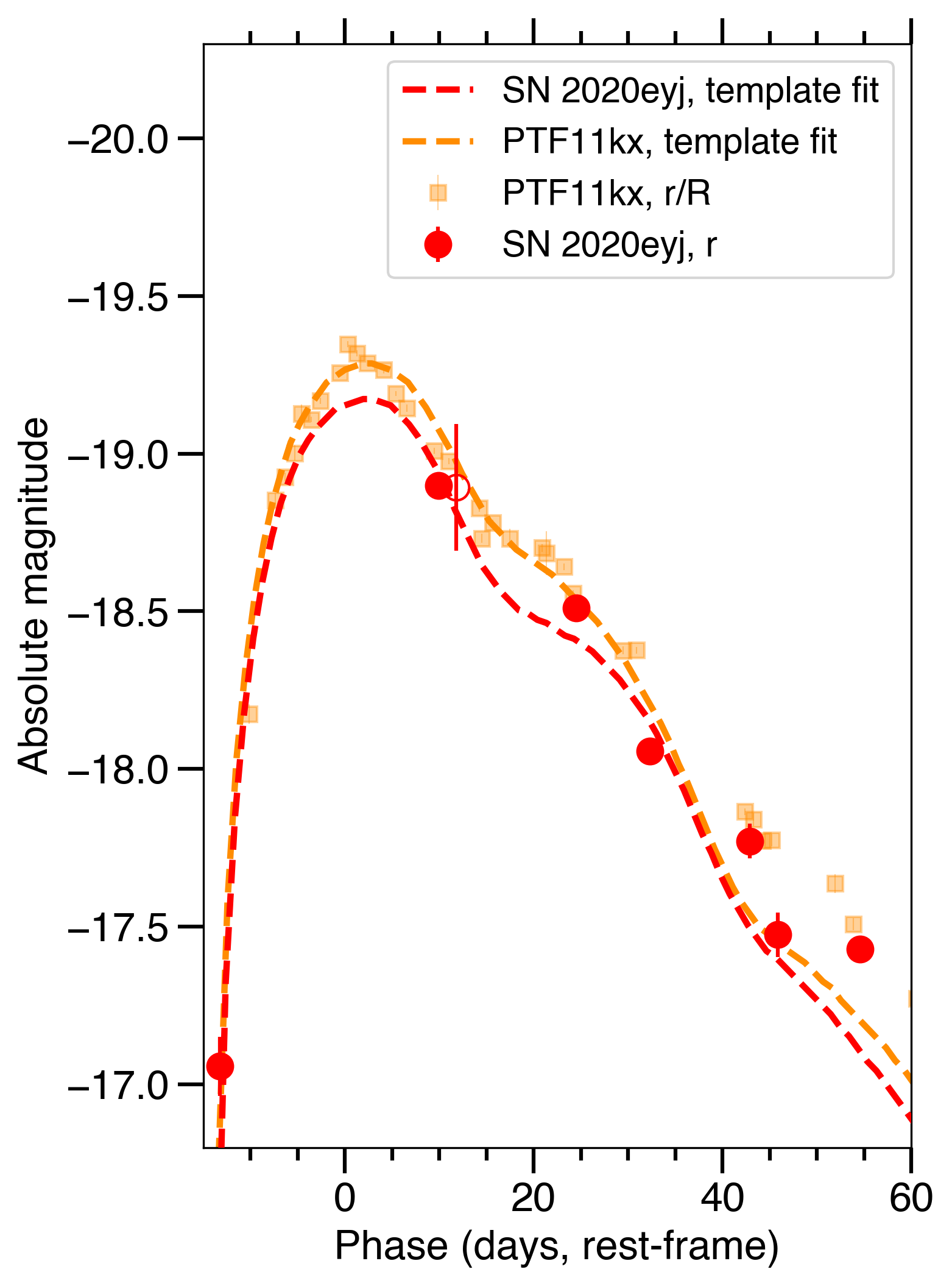}
         \caption{$r$ band}
     \end{subfigure}
    \begin{subfigure}[t]{0.32\textwidth}
         \centering
         \includegraphics[width=0.99\textwidth]{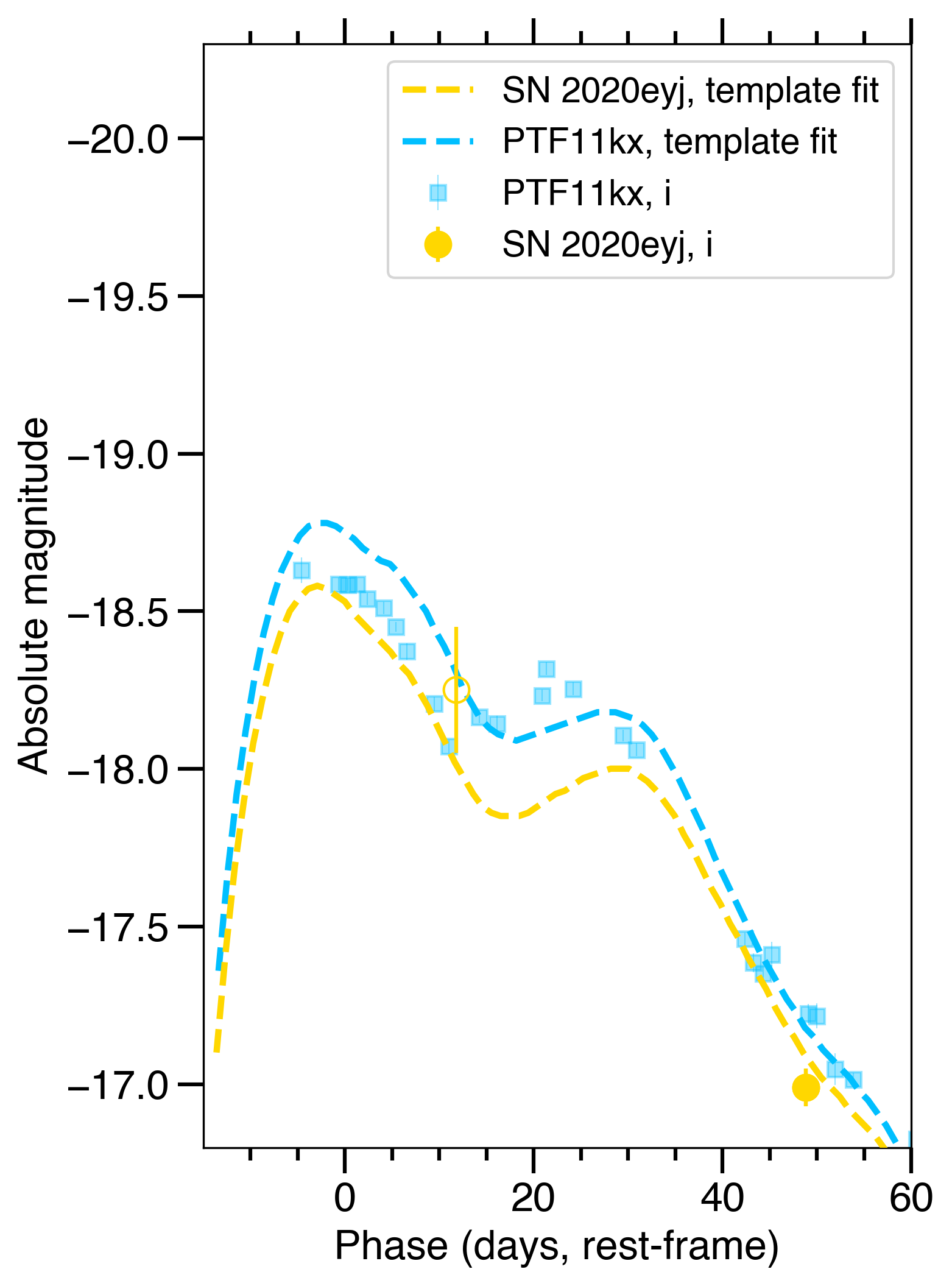}
         \caption{$i$ band}
     \end{subfigure}
\caption{{\bf The light curves of \eyj are consistent with a SN Ia and its H-rich analog SN~Ia-CSM PTF11kx}. We simultaneously fit the $g$, $r$ and $i$ light curves of the initial peak phases of both \eyj and PTF11kx with the SN Ia light curve fitter SNooPy. \eyj is well fit with stretch factor 1.2 and E(B$-$V) =  0.5 $\pm$ 0.1 mag. Similarly, PTF11kx is well fit with stretch factor 1.2 and E(B$-$V) =  0.25 $\pm$ 0.02 mag. Panels show the absolute magnitude light curves of \eyj and PTF11kx, after correcting for the host extinction derived from the fit. (a) $g$ band; (b) $r$ band for \eyj and $r/R$ band for PTF11kx; (c) $i$ band. Open circles indicate synthetic photometry derived from the spectra. The error bars represent 1$\sigma$ uncertainties.} 
\label{fig:lightcurve_fits} 
\end{figure*}

\clearpage

\begin{figure*}
\centering
    \begin{subfigure}[t]{0.49\textwidth}
         \centering
         \includegraphics[width=0.7\textwidth]{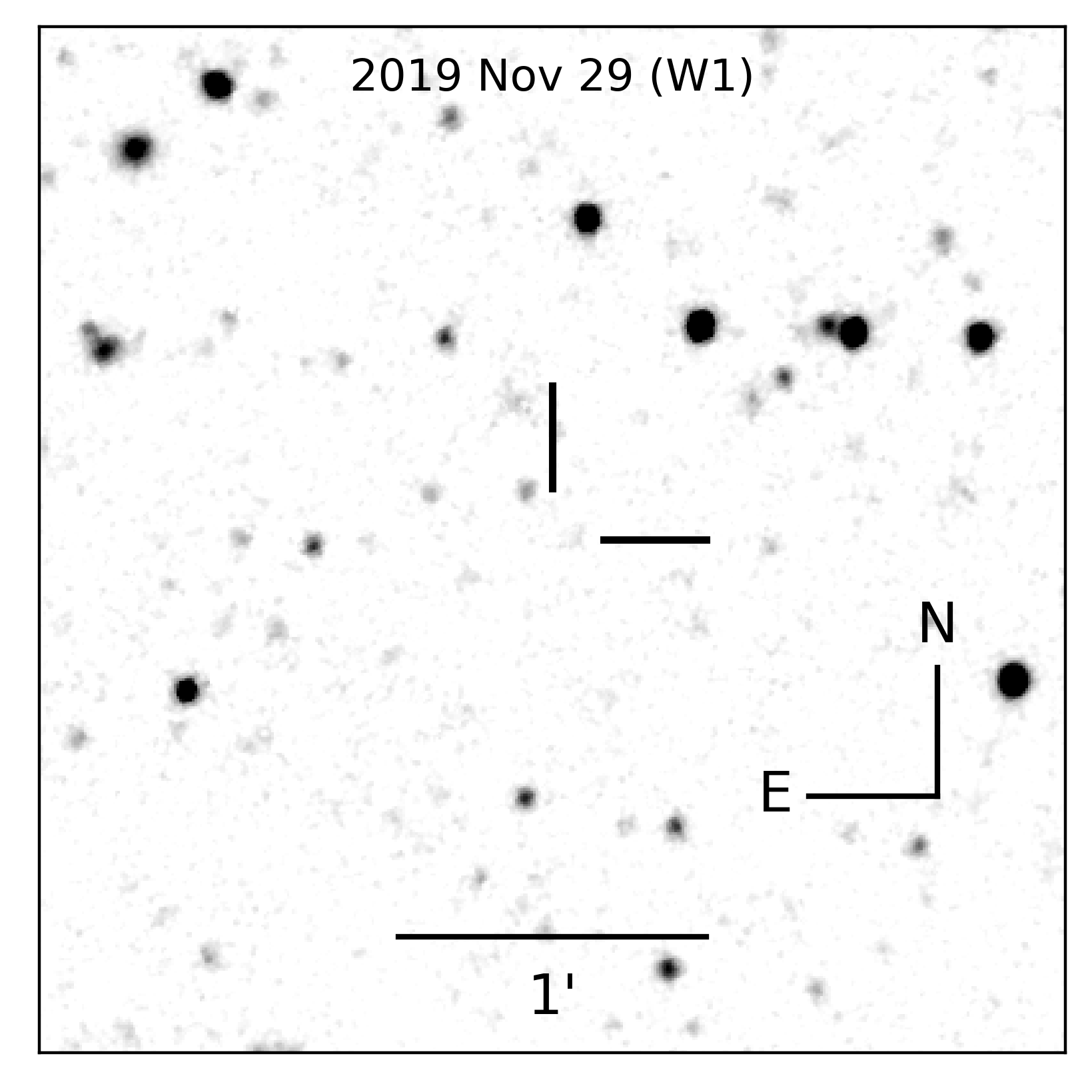}
         \caption{Mid-IR pre-explosion}
     \end{subfigure}
    \begin{subfigure}[t]{0.49\textwidth}
         \centering
         \includegraphics[width=0.7\textwidth]{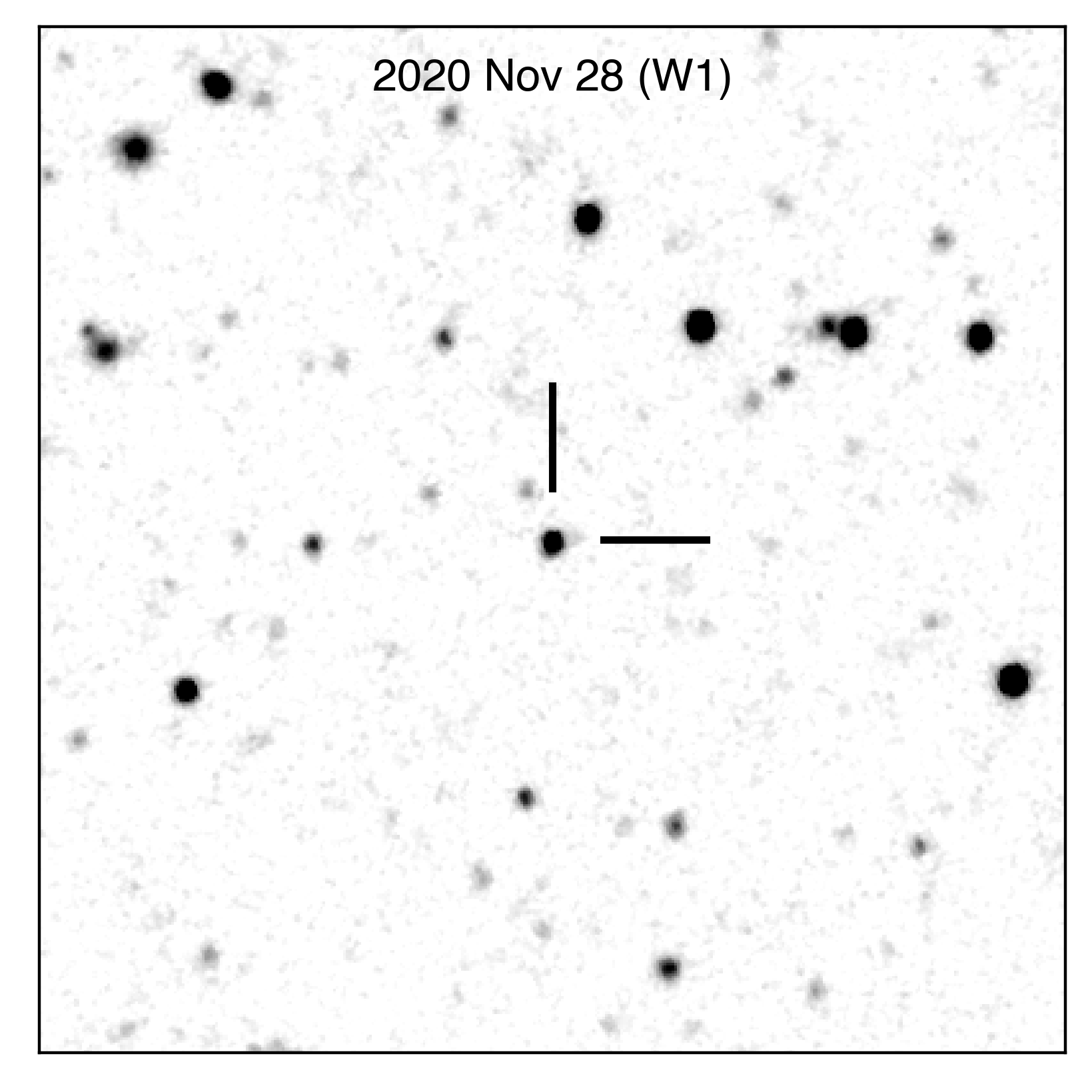}
         \caption{Mid-IR detection of \eyj}
     \end{subfigure}
    \begin{subfigure}[b]{0.99\textwidth}
         \centering
         \includegraphics[width=0.7\textwidth]{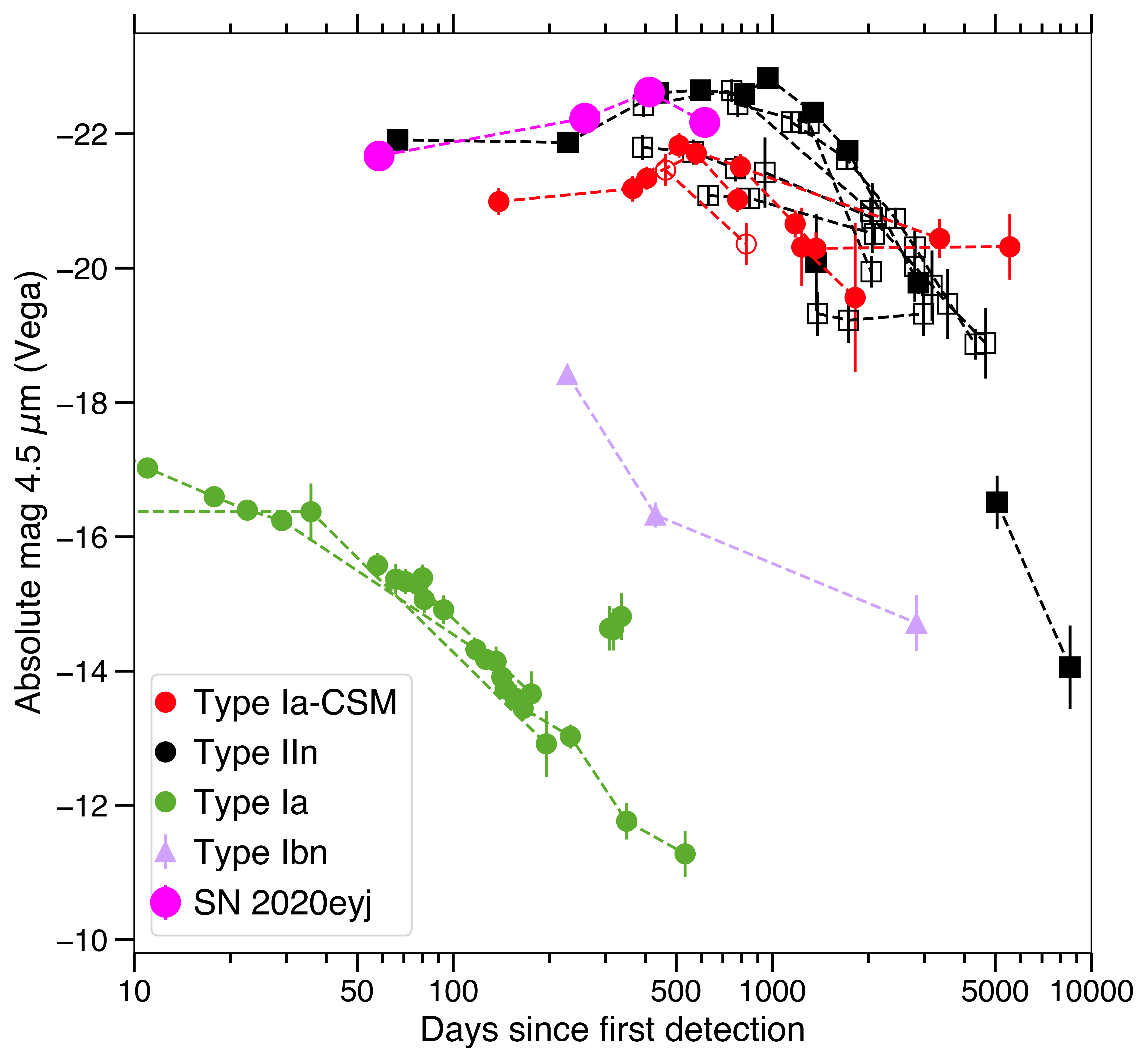}
         \caption{Mid-IR light-curve comparison}
     \end{subfigure}
\caption{{\bf \eyj was accompanied by a bright mid-IR counterpart}. (a) A coadded image of the last NEOWISE-R epoch before the SN explosion, without any sign of the SN host. (b) The coadded image in the $W1$ filter of the 2020 November NEOWISE-R epoch, 261 days after first detection, with \eyj clearly visible. (c) A mid-IR light curve comparison of \eyj in the $W2$ filter (4.6 $\mu$m) to a sample of SNe observed with Spitzer at 4.5$\mu$m, adapted from \cite{Szalai2021}, including Type IIn SNe (in black), Type Ia-CSM SNe (in red), and Type Ibn SN 2006jc (in lilac). Additionally, the light curves of a sample of SNe Ia from \cite{Johansson2017} is plotted in green. \eyj (in pink large circles) is among the brightest SNe observed in the mid-IR, and is 6$-$10 magnitudes brighter than normal SNe Ia, depending on the phase. The error bars represent 1$\sigma$ uncertainties. } 
\label{fig:wise} 
\end{figure*}

\clearpage

\begin{figure}[t!]
  \centering
  \includegraphics[width=0.99\columnwidth]{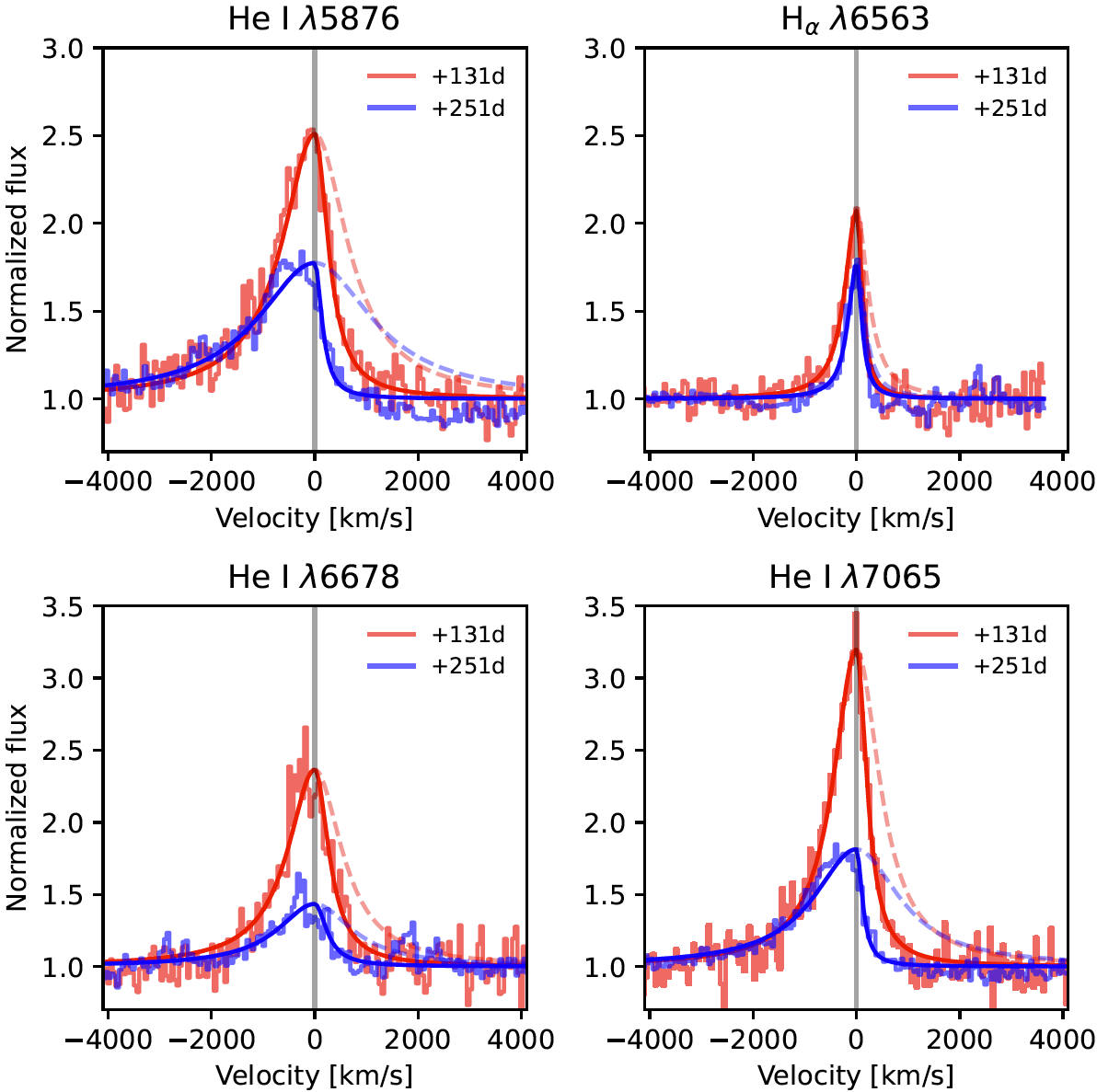}
  \caption{{\bf The He and H$\alpha$ emission line profiles in the late spectra of \eyj show significant asymmetry.} The He\,{\sc i} emission lines at 5876 \AA, 6678 \AA\ and 7065 \AA\ all show strong attenuation in the red wings, and an apparent blue shift over time between the 131 and 251 days epochs. Such line asymmetry is commonly observed in SNe Ia-CSM \cite{Silverman2013}, and is interpreted as due to the condensation of dust in the ejecta or shocked CSM, obscuring the red wing (Sect.~\ref{sec:infrared}), but may also be a result of optical depth effects \cite{Dessart2022}. The H$\alpha$ emission line at 131 days also shows asymmetry and there is a (minor) decline in flux between the two epochs shown here. By 329 days, the H$\alpha$ luminosity has dropped to the level of the line emission in the host spectrum (Sect.~\ref{sec:spectroscopy}).}
  \label{fig:emlines} 
\end{figure}



\clearpage

\begin{table}
\caption{\label{tab:spec}Log of spectroscopic observations of \eyj and its host, and FWHM velocity measurements of prominent emission lines in the 131 and 251 days Keck spectra. Phase is relative to first detection epoch, in rest-frame. The listed velocities are both the FWHM velocities measured from fitting the full line with a Lorentzian line profile, as well as twice the half width at half maximum of the blue wing. The latter measurements better represents the true FWHM velocity, because the red wings in the emission lines are strongly attenuated (Fig.~\ref{fig:emlines}).} 
\centering
\begin{tabular}{lllll} \hline
MJD     && Date  &Phase   &Telescope + Instrument \\
        &&  (UT)                 &(rest-frame)       & \\ \hline
58941.2 && 2020 Apr 02       & 25                    & P60+SEDM \\
59050.3 && 2020 Jul 20       & 131                   & Keck1+LRIS  \\
59150.5 && 2020 Oct 28       & 228                   & P60+SEDM  \\
59162.2 && 2020 Nov 09       & 240                   & NOT+ALFOSC  \\
59173.6 && 2020 Nov 30       & 251                   & Keck1+LRIS  \\
59253.9 && 2021 Feb 08       & 329                   & NOT+ALFOSC  \\
59615.4 && 2022 Feb 05       & 678                   & Keck1+LRIS  \\ 
\hhline{=====}
\multicolumn{5}{c}{Line velocities}\\ \hline
            &He\,{\sc i} $\lambda 5876$     &He\,{\sc i} $\lambda 6678$     &He\,{\sc i} $\lambda 7065$     &H$\alpha$ \\
&(km s$^{-1}$)&(km s$^{-1}$)&(km s$^{-1}$)&(km s$^{-1}$) \\ \hline
\textbf{131 days}&&&&\\\hline
Full        &1080                               &960                                &780                                &400\\
Blue wing   &1540                               &1270                               &1080                               &540\\ \hline
\textbf{251 days}&&&&\\\hline
Full        &1500                               &1150                               &1140                               &320\\
Blue wing   &2680                               &1750                               &2000                               &390\\ \hline
\end{tabular}
\end{table}


\clearpage

\begin{table}
\caption{\label{tab:wise}Mid-IR photometry from the WISE telescope and infrared and dust properties of \eyj. Phase (row 2, rest-frame days since first detection), magnitudes (rows 3 and 4, in AB system and binned per epoch), dust mass and temperature (rows 5 and 6, assuming 0.1$\mu$m amorphous carbon grains), blackbody temperature, radius and luminosity (rows 7 - 9), and the cumulative radiated energy (row 10).} 
\centering
\begin{tabular}{llllll} \hline
                                            && Epoch 1 &  Epoch 2   & Epoch 3 & Epoch 4  \\\hline
MJD&                                         & 58975.45 &  59181.42   & 59339.60 & 59548.01 \\
Phase&(rest-frame days)                      & 58.5 &  258.5   & 412.0 & 614.4 \\
W1& (mag)                                    & 17.29 $\pm$ 0.03 &  16.72 $\pm$ 0.02   & 16.40 $\pm$ 0.02 & 17.30 $\pm$ 0.04 \\
W2& (mag)                                    & 17.26 $\pm$ 0.03 &  16.70 $\pm$ 0.04   & 16.31 $\pm$ 0.02 & 16.76 $\pm$ 0.03 \\
$M_{\textrm{dust}}$& (10$^{-3}$ \msun)       & 1.8 $\pm$ 0.3 &  2.8 $\pm$ 0.5   & 4.8 $\pm$ 0.5 & 9.9 $\pm$ 2.1 \\
$T_{\textrm{dust}}$& (K)                     & 801 $\pm$ 23 &  809 $\pm$ 27   & 778 $\pm$ 16 & 608 $\pm$ 23 \\
$T_{\textrm{BB}}$& (K)                       & 1268 $\pm$ 63 &  1291 $\pm$ 27   & 1201 $\pm$ 38 & 826 $\pm$ 35 \\
$r_{\textrm{BB}}$& (10$^{16}$ cm)            & 2.5 $\pm$ 0.2 &  3.2 $\pm$ 0.3   & 4.2 $\pm$ 0.2 & 6.4 $\pm$ 0.6 \\
$L_{\textrm{BB}}$& (10$^{42}$ erg s$^{-1}$)  & 1.2 $\pm$ 0.3 &  2.0 $\pm$ 0.4   & 2.6 $\pm$ 0.4 & 1.4 $\pm$ 0.3 \\
Cumulative& (10$^{49}$ erg)                  & - &  2.7 $\pm$ 0.4   & 5.8 $\pm$ 0.7 & 9.3 $\pm$ 1.0 \\ \hline
\end{tabular}
\end{table}

\clearpage
 
\begin{figure*}
\centering
    \begin{subfigure}{0.49\textwidth}
         \centering
         \includegraphics[width=\textwidth]{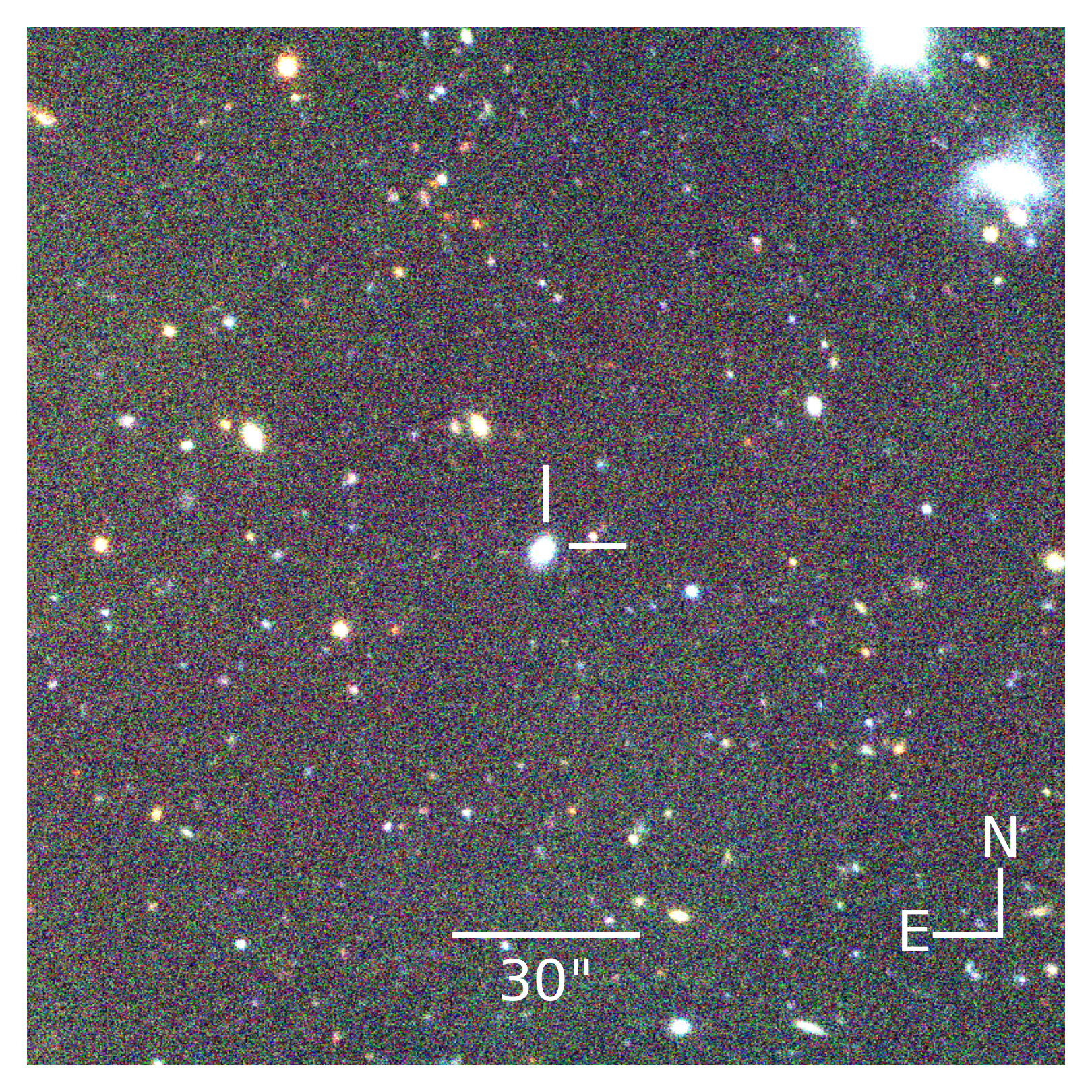}
         \caption{Host and environment}
     \end{subfigure}
    \begin{subfigure}{0.49\textwidth}
         \centering
         \includegraphics[width=\textwidth]{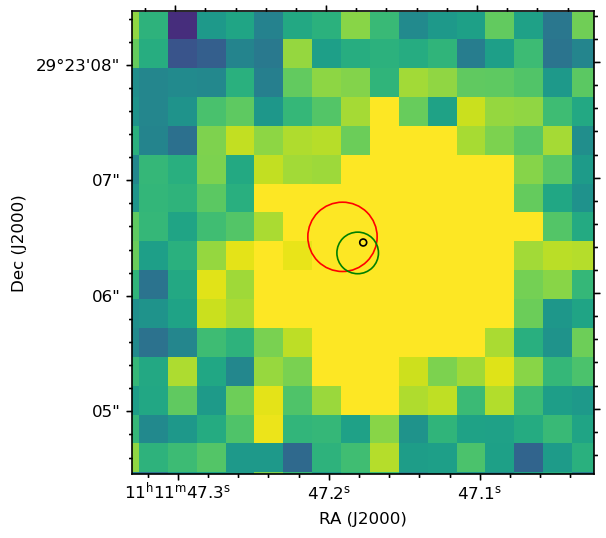}
         \caption{Close--up of host}
     \end{subfigure}
\caption{ {\bf The position of the radio detection is consistent with the position of \eyj in the optical}. (a) A $3^{\prime}\times3^{\prime}$ color composite image, obtained with NOT/ALFOSC, of the compact star-forming host galaxy of \eyj and its environment. (b) The average position of the e-MERLIN detections (black circle, 0.01$^{\prime\prime}$ uncertainty), the position reported in GaiaAlerts ($G$ band, green circle, 0.06$^{\prime\prime}$ uncertainty), and the position of \eyj in the ALFOSC epoch at 382 days ($r$ band, red circle, 0.1$^{\prime\prime}$ uncertainty), overlaid on a $4^{\prime\prime}\times4^{\prime\prime}$ Pan-STARRS1 $i$-band data of the host.} 
\label{fig:astrometry}
\end{figure*}

\clearpage

\begin{figure}[t!] 
  \centering
  \includegraphics[width=0.75\columnwidth]{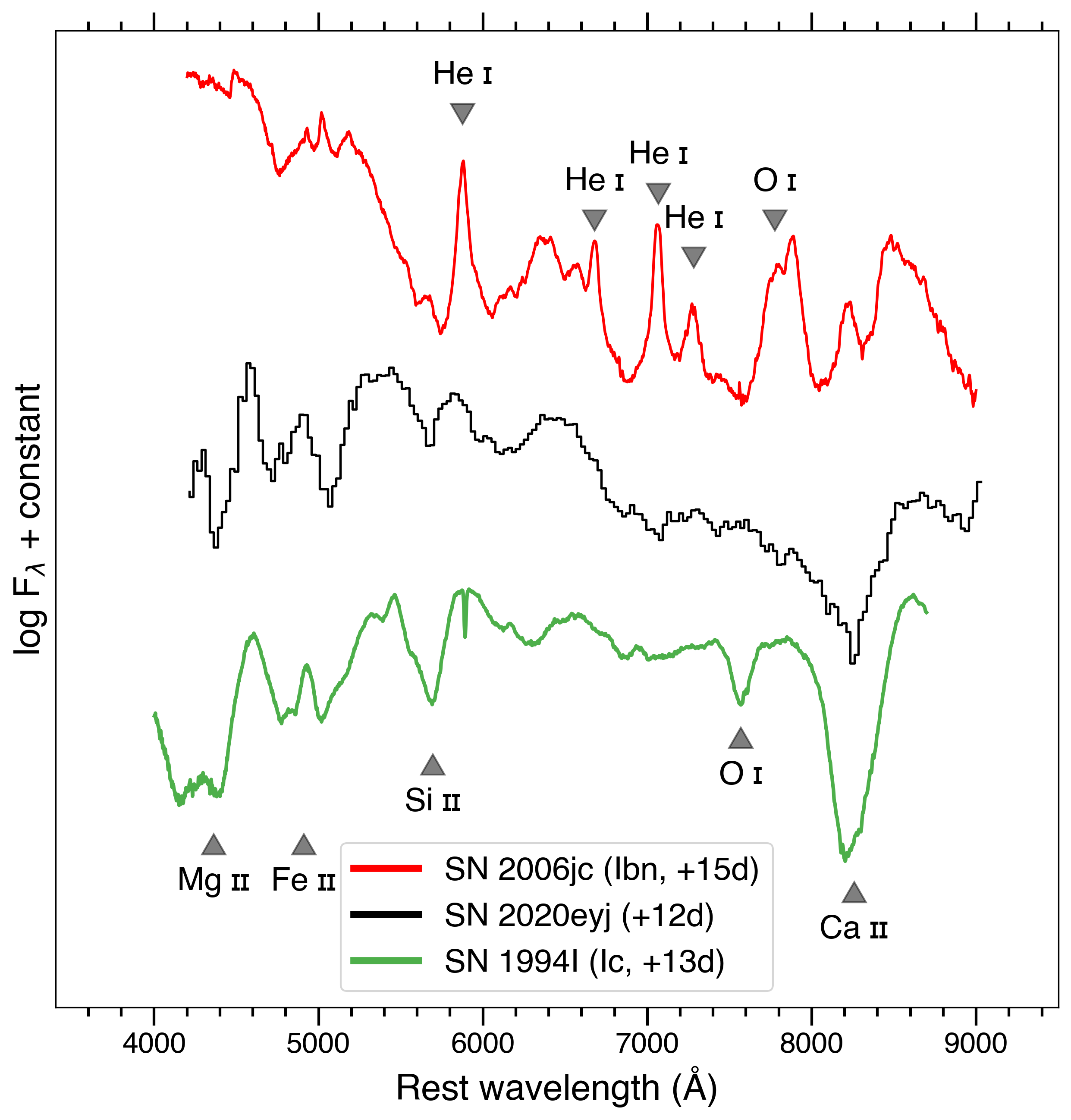}
  \caption{{\bf The spectra of \eyj do not match Type Ic or Ibn SNe at early epochs.} The SEDM classification spectrum of \eyj compared with Type Ibn SN~2006jc and Type Ic SN1994I at similar epochs, $\sim$12 days after peak. At 15 days post peak, SN~2006jc already showed strong He emission lines and developed the quasi-continuum typical for CSM-interaction dominated spectra. These features are not observed in \eyj at 12 days post peak, but do become prominent at late phases (Fig.~\ref{fig:laterspectra}). At 13 days post peak, SN~1994I has grown redder compared to its peak spectrum shown in Fig.~\ref{fig:classificationspectrum}, and the O\,{\sc i} 7774 \AA\ absorption feature has become more prominent, whereas in \eyj this feature is not present.}
  \label{fig:classificationspectrum_methods} 
\end{figure}

\clearpage

\begin{figure}[t!]
  \centering
  \includegraphics[width=0.99\columnwidth]{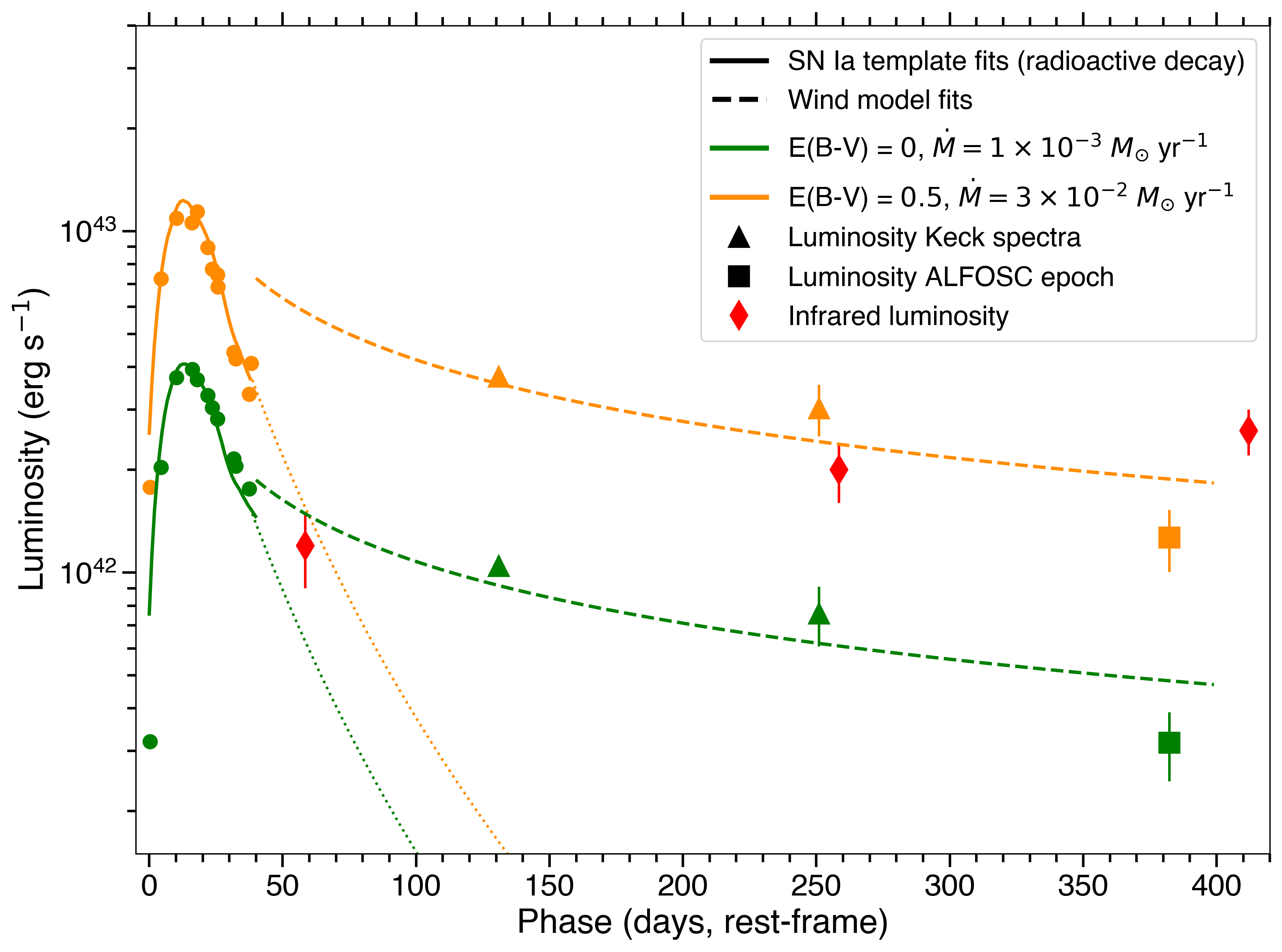} 
  \caption{{\bf The bolometric light curve of \eyj can be described with a radioactive decay model for the peak phase, and an optically thick wind for the tail phase.} For the initial SN Ia peak of \eyj we adopt the bolometric light curve (solid lines) accompanying the SN Ia template fit to the $gri$ photometry (Sect.~\ref{sec:lightcurve}), assuming no line of sight extinction (in green) and an extinction of E(B-V) = 0.5 mag (in orange). Overplotted are the associated bolometrically corrected luminosities up until 40 days. From epoch 46 days onward the SN Ia template fit does not accurately describe the observed ($g$ band) photometry any longer (Fig.~\ref{fig:lightcurve_fits}). The dotted lines show the continuation of the bolometric light curve of the underlying SN Ia. The three measurements in the tail phase are based on the integration of the two Keck spectra, extrapolated to the UV, and a bolometrically corrected photometric ALFOSC epoch. The striped lines represent the fits to the tail phase measurements using the analytical model from \cite{Moriya2019}, following the same color scheme for the level of extinction. In the transition region from the diffusion peak to the CSM interaction-powered tail, between 50 and 100 days, the sum of the models would overestimate the luminosity, suggesting the CSM configuration is more complicated than a simple wind-like density profile. The red diamonds show the IR luminosity of \eyj (Table~\ref{tab:wise}), which are not included in the model fits.}
  \label{fig:model_comparison} 
\end{figure}

\clearpage

\begin{table}
	\caption{\label{tab:hostphot} Host galaxy photometry. Magnitudes are not corrected for reddening.} 
	\centering

	\begin{tabular}{cccc} \hline
	Survey/Telescope & Instrument & Filter & Magnitude \\ \hline
	GALEX			 & 				&$FUV	$&$21.37\pm0.38$\\
	GALEX			 & 				&$NUV	$&$20.99\pm0.14$\\
	Swift			 & UVOT 		&$w2	$&$21.45\pm0.16$\\
	Swift			 & UVOT 		&$m2	$&$21.66\pm0.27$\\
	Swift			 & UVOT 		&$w1	$&$21.31\pm0.24$\\
	SDSS			 & 				&$u		$&$21.16\pm0.23$\\
	SDSS			 & 				&$g		$&$20.26\pm0.06$\\
	SDSS			 & 				&$r		$&$19.86\pm0.05$\\
	SDSS			 & 				&$i		$&$19.74\pm0.13$\\
	PanSTARRS		 & 				&$g		$&$20.46\pm0.10$\\
	PanSTARRS		 & 				&$i		$&$19.94\pm0.08$\\
    PanSTARRS		 & 				&$z		$&$19.71\pm0.16$\\
	PanSTARRS		 & 				&$y		$&$20.21\pm0.57$\\
	NOT				 & ALFOSC		&$r		$&$19.99\pm0.03$\\
	\hline
	\end{tabular}
\end{table}

\clearpage


\end{extended_data}

\newpage

\begin{methods}

\section{Observations}

\subsection{Discovery}\label{sec:discovery}
\eyj was discovered by the Asteroid Terrestrial-impact Last Alert System (ATLAS) \cite{tonry2018,smith2020} on 2020 March 23 UT \cite{tonry2020}, and subsequently detected as part of the Zwicky Transient Facility (ZTF) survey \cite{Bellm2019,graham2019b}, at $\alpha=11^{h}11^{m}47.19^{s}$, $\delta=$ +29\textdegree23$^{\prime}$06.5$^{\prime\prime}$ (J2000). Pre-discovery detections were recovered in the ZTF data on 2020 March 07 UT (MJD = 58915.12) in both $g$ and $r$ filters. For reference, we list some key characteristics of \eyj in Table~\ref{tab:properties}.

\subsection{Optical photometry}\label{sec:photometry}

Follow-up photometry was obtained as part of public and partnership ZTF survey observations \cite{bellm2019b} with the ZTF camera \cite{dekany2020} on the P48 telescope in the $g$ and $r$ bands, and later phases were also covered in the $i$ band. The P48 data were reduced and host subtracted using the ZTF reduction and image subtraction pipeline \cite{masci2019}, which makes use of the ZOGY algorithm \cite{Zackay2016} for reference image subtraction. Following the rationale illustrated in \cite{Yuhan2019}, we apply the difference image zero point magnitude to convert fluxes from units in detector data number (DN) to $\mu$Jy, and translate fluxes to AB magnitudes. We apply a detection threshold of S/N $\gtrsim$ 3, and for non-detections we compute 5 sigma upper limits. Table~\ref{tab:photometry} lists the ZTF magnitudes and upper limits.

Additional photometric epochs were obtained with the Liverpool Telescope (LT) \cite{steele2004}, the SEDM on the P60, the LCO telescopes (program id. NOAO2020B-012), and ALFOSC on the NOT, with data reduced and host subtracted using the pipelines described in \cite{fremling2016,De2020} or standard methods. In this work we also make use of the forced photometry service from the ATLAS survey \cite{tonry2018,2020PASP..132h5002S}, which contained valuable photometry in the $o$ and $c$ bands. One $i$-band epoch was obtained from the Pan-STARRS1 data archive \cite{Chambers2016}.

ZTF and ATLAS also obtained observations of the location of \eyj in the nights immediately preceding the first detection, with limiting magnitudes in the ZTF $g$ band on 2020 March 5 and 6 UT of 20.8 and 19.7, respectively, and the (binned) observations in $o$ band from ATLAS on March 5 UT correspond to a limiting magnitude of 20.2. Phases in this study are relative to the first ZTF detection (MJD = 58915.212, 2020 March 7 UT) in rest-frame days, unless stated otherwise. Given the excellent constraints on the nights before, this epoch is also close to the time of first light.

All magnitudes are reported in the AB system. The extinction in the Milky Way (MW) was obtained from \cite{Schlafly2011} as E(B$-$V) = 0.024 mag. MW reddening corrections are applied using the \cite{cardelli1989} extinction law with $R_V=3.1$, whereas SN reddening corrections are applied using $R_V=2$. The photometric magnitudes of \eyj are listed in Table~\ref{tab:photometry}. The ATLAS and P48 light curves are shown in Fig.~\ref{fig:lightcurve}, binned into 1-night bins to enhance the signal to noise ratio (S/N).

\subsection{Optical spectroscopy}\label{sec:spectroscopy}

The first optical spectrum of \eyj was obtained with the Spectral Energy Distribution Machine (SEDM) \cite{blagorodnova2018} mounted on the Palomar 60-inch telescope (P60) \cite{cenko2006}, 25 days after first detection. All SEDM spectra are automatically reduced and calibrated with \texttt{pysedm} \cite{rigault2019}, and the quality of the SEDM spectrum of \eyj was further improved using \texttt{hypergal} \cite{Lezmy2022}. Follow-up spectroscopy was obtained from 131 days onward with the Low-Resolution Imaging Spectrometer (LRIS) \cite{Oke+1995} on the Keck I telescope, and the Alhambra Faint Object Spectrograph and Camera (ALFOSC) on the Nordic Optical Telescope (NOT) \cite{djupvik2010}. A host spectrum was obtained at 678 days, after the SN had fully faded from view. The spectra were reduced in a standard manner, using \textsc{lpipe} \cite{Perley2019a} and \textsc{PypeIt} \cite{Prochaska2020,Prochaska2020b} for Keck/LRIS and NOT/ALFOSC, respectively.

A log of the obtained spectra is provided in Table~\ref{tab:spec}, and the epochs of spectroscopy are indicated by the diamond markers on top of the light curves in Fig.~\ref{fig:lightcurve}. The spectra were absolute flux-calibrated against the $r$-band magnitudes using the Gaussian Process interpolated magnitudes and then corrected for MW extinction. All spectral data and corresponding information will be made available via WISeREP public database \cite{wiserep}. We present the peak SEDM spectrum in Fig.~\ref{fig:classificationspectrum} and the later sequence of spectra in Fig.~\ref{fig:laterspectra}. 

The initial spectrum obtained with SEDM is characterized by broad absorption features (Sect.~\ref{sec:snia}). The later spectra are shaped by broad Fe\,{\sc ii} lines, in particular the quasi-continuum blueward of 5700 \AA \cite{chugai2009,Kiewe2012,smith2012,Stritzinger2012,pastorello2015IV}, and a prominent Ca\,{\sc ii} near-IR triplet. Superimposed on the continuum are narrow He\,{\sc i} emission lines, as well as H$\alpha$. We measure FWHM velocities of the He\,{\sc i} emission lines and H$\alpha$ in the spectra obtained with Keck at 131 and 251 days, by fitting a Lorentzian profile to the complete lines, as well as to just the blue wings. The red wings in the He and H$\alpha$ lines are significantly attenuated (Fig.~\ref{fig:emlines} and Sect.~\ref{sec:infrared}), so the intrinsic FWHM velocities are better represented by (double) the blue wing FWHM. We report these FWHM velocities in Table~\ref{tab:spec}. The FWHM velocities of the He\,{\sc i} emission lines range from 1100 to 2700 km s$^{-1}$ (corrected for the red wings), with no sign of a narrow ($<1000$ km$^{-1}$) component detected in some SNe Ibn and interpreted as coming from the unshocked CSM \cite{pastorello2008II,smith2012}. There is also no sign of material stripped from the donor star by the SN ejecta \cite{Botyanszki2018,Dessart2020}, which is predicted to show up as narrow emission ($<1000$ km$^{-1}$ \cite{Pan2010}).

The asymmetric line profile we associate with the SN also applies to the H$\alpha$ emission line, suggesting the presence of H in the CSM. In the spectrum obtained at 131 days, H$\alpha$ has an equivalent width (EW) of 14~\AA, not corrected for contribution by the host. By comparison, the He\,{\sc i} emission lines at 5876, 6678 and 7065~\AA\ in the same spectrum have EWs of 47~\AA, 43~\AA\ and 61~\AA, respectively. As H is easier to ionize than He, the more prominent He lines means that the CSM must predominantly consist of He. By epoch 329 days, the H$\alpha$ luminosity has dropped to the luminosity of the H$\alpha$ narrow emission line in the host spectrum obtained at 678 days (Sect.~\ref{sec:host}). 

\subsection{Infrared photometry}

Following a report \cite{thevenot2021} of a mid-IR detection of \eyj in the 2021 data release of the NEOWISE Reactivation (NEOWISE-R) \cite{mainzer2014} survey, we queried the IPAC Infrared Science Archive for any NEOWISE-R detections at the position of \eyj. After filtering out poor quality data and binning individual exposures following the method described in \cite{kool2020a}, the SN was recovered in both $W1$ and $W2$ filters (3.4 and 4.6 $\mu$m, respectively) in all four 2020 and 2021 epochs, with the earliest detection at 59 days after first detection (Fig.~\ref{fig:lightcurve} and Table~\ref{tab:wise}). The host is not detected in (stacked) WISE data prior to the SN explosion (Fig.~\ref{fig:wise}, top panels), so we assume the contribution from the host is negligible and all observed flux is due to the SN.

\subsection{Radio}\label{sec:radio}

We observed \eyj with the electronic Multi-Element Radio Linked Interferometre Network (e-MERLIN) in two epochs. The first epoch, with a duration on target and phase calibrator of $\sim$16 hours, was conducted on 2021 November 19 (centred on MJD 59538.29), 605 days after first detection and included six e-MERLIN telescopes (Mk2, Kn, De, Cm, Da and Pi). The second epoch was conducted during 6 consecutive days between 2022 April 6 and 12 (mean MJD 59678.59, 741 days after first detection). Between 5 and 6 telescopes (including the Lovell) participated, with some antennae missing part of the runs due to technical problems. Due to the significantly smaller field of view of the Lovell telescope, the pointing centre of the second epoch was shifted by 1 arcmin to include an inbeam calibrator in the primary beam of this telescope. 3C 286 and OQ 208 were used as amplitude and bandpass calibrators, respectively. The phase calibrator, J1106+2812, was correlated at position $\alpha_{\rm J2000.0}=11^{\rm h} 06^{\rm m} 07.2617^{\rm s}$ 
and $\delta_{\rm J2000.0}=$ 28\textdegree 12' 47.065'', at a separation of 1.7\textdegree\ from the target and was detected with a flux density of 150~mJy. We centered our observations at a frequency of 5.1 GHz, using a bandwidth of 512 MHz. The data were correlated with the e-MERLIN correlator at Jodrell Bank Observatory (JBO), using 4 spectral windows, each of 512 channels, with 1 sec integrations and 4 polarizations. 

We calibrated and processed the data using the e-MERLIN CASA pipeline \cite{moldon21} version v1.1.19 running on CASA version 5.6.2. We used the 10 mJy inbeam source to self-calibrate the residual phases and amplitudes of the target source. Cleaning was done with the software package \texttt{wsclean} \cite{offringa-wsclean-2014}. Final images of the target were produced with a synthesized beam of 80~mas $\times$ 35~mas at a P.A. of 28\textdegree, and 94~mas $\times$ 71~mas at a P.A. of $-71$\textdegree, in the first and second epoch, respectively. The 1-$\sigma$ rms of the images is 17 and 8~$\mu$Jy beam$^{-1}$, respectively. The target is detected in both epochs as an unresolved source as characterized with task IMFIT. We estimate the uncertainty of the peak flux density to be a quadratic sum of the image rms and a conservative 10\% amplitude scale calibration error. The final flux density of the source is $80\pm20$ and $60\pm10$~$\mu$Jy beam$^{-1}$ in the first and second epoch, respectively. The radio source is located at an average position of $\alpha_{\rm J2000.0}=11^{\rm h} 11^{\rm m} 47.1763^{\rm s}$ and $\delta_{\rm J2000.0} = $29\textdegree 23'06.45'', with an estimated uncertainty of 10~mas.

The average position of the e-MERLIN detections relative to the optical positions of \eyj is shown in Fig.~\ref{fig:astrometry}. The radio detection is consistent with the position of the SN in the ALFOSC epoch at 382 days ($r$ band), and the position reported in GaiaAlerts of the detection of \eyj in $G$ band at 42 days.

\subsection{X-ray}\label{sec:xray}

On 27 April 2022, 758 days after first detection, we observed \eyj for 3.8~ks with the X-ray telescope XRT between 0.3 and 10 keV aboard the Neil Gehrels Swift Observatory \cite{Gehrels2004a, Burrows2005a}. We analyzed the data with the online-tools of the UK \textit{Swift} team (\url{https://www.swift.ac.uk/user_objects/}) that use the methods described in \cite{Evans2007a, Evans2009a} and the software package \texttt{HEASoft} version 6.26.1. \eyj evaded detection down to a count-rate of 0.003~count~s$^{-1}$ ($3\sigma$ limit). To convert the count-rate limit into a flux limit, we assumed a power-law spectrum with a photon index $\Gamma$ of 2 and a Galactic neutral hydrogen column density of $1.9\times10^{20}$~cm$^{-2}$ \cite{HI4PI2016a}. Here the photon index $\Gamma$ is defined as the power-law index of the photon flux density ($N(E)\propto E^{-\Gamma}$). Between 0.3--10 keV the count-rate limit corresponds to an unabsorbed flux of $1.1\times10^{-13}~{\rm erg\,cm}^{-2}\,{\rm s}^{-1}$ and a luminosity of $<2.4\times10^{41}~{\rm erg}\,{\rm s}^{-1}$. It is possible a deeper observation would have yielded a detection, as the Type Ia-CSM SN~2012ca was detected in X-rays at a similar epoch, with a luminosity of order $10^{40}~{\rm erg}\,{\rm s}^{-1}$ \cite{Bochenek2018}.

\section{SN Ia classification}\label{sec:snia}

During the peak phase of \eyj, an optical spectrum was obtained with the low-resolution (R$\sim$100) SEDM on the P60, 25 days after first detection. This high S/N spectrum was characterized by broad absorption features (Fig.~\ref{fig:classificationspectrum}), based on which \eyj was classified as a Type Ia SN at redshift $z=0.03$ \cite{Dahiwale2020}. Using \texttt{SNIascore}, a deep-learning-based classifier of SNe Ia based on low-resolution spectra, \cite{Fremling2021} noted that the SN could be a Type Ibc SN erroneously classified as SN Ia due to the degeneracy between peak spectra of SNe Ibc with those of SNe Ia at post-peak phases, but their classifier anyway favored a SN Ia classification. In general, based on the comparison study by \cite{Leloudas2015}, Type Ibc SNe erroneously classified as Type Ia(-CSM/91T) are a lot less common than the inverse. Here we scrutinize the SEDM spectrum using comparisons with SNe from the literature, based on spectral matching with the \texttt{SuperNova} \texttt{IDentification} (\texttt{SNID}) \cite{blondin2007} and \texttt{Superfit} \cite{Howell2005} classification tools, where the \texttt{SNID} template library has been supplemented with the Type Ibc templates from \cite{Modjaz2014}. We adopt a $g$-band peak epoch of MJD = 58929 $\pm$ 2, based on the light curve fitting described in Sect.~\ref{sec:lightcurve}, with the uncertainty driven by the poor sampling of our photometry around peak.

The top 10 \texttt{SNID} (\texttt{rlap} $>$ 10) and \texttt{Superfit} matches are all of Type Ia (Fig.~\ref{fig:classificationspectrum}), and include normal SNe Ia such as SN 2004eo \cite{Pastorello2007a} and 91T-like SNe such as SN 2001V \cite{Vinko2003}. The best matching SN of Type Ibc (\texttt{rlap} = 8) is the Type Ic SN 1994I \cite{Nomoto1994,Sasaki1994,Wheeler1994,Filippenko1995}. The phases corresponding to the matched SNe Ia are all post-peak, ranging from 12 days to $\sim$50 days post-maximum, whereas the matching SN Ibc spectra are all within a few days from peak. The phase of the SEDM spectrum of \eyj is 12 days post-maximum, which corroborates the SN Ia classification.

In terms of spectral features, the SEDM spectrum shows broad absorption lines that based on the spectral comparisons can be unambiguously identified as Si\,{\sc ii}, Fe\,{\sc ii} and Ca\,{\sc ii} (Fig.~\ref{fig:classificationspectrum}). Compared to normal SNe Ia as exemplified by SN 2004eo, the Si\,{\sc ii} features in \eyj are quite shallow. Diluted Si\,{\sc ii} absorption is common for 91T-like SNe Ia, as in the spectrum of SN 2001V. Type Ia-CSM are known to show 91T-like spectra around peak \cite{Leloudas2015}. As a SN strongly interacting with a CSM, the presence of diluted Si\,{\sc ii} in the SEDM spectrum of \eyj is consistent with a Type Ia(-CSM) classification. In terms of expansion velocity, the velocity of the Si\,{\sc ii} $\lambda$6355 absorption feature in the SEDM spectrum is $8900 \pm 600$ km s$^{-1}$. This velocity is on the slow side for the SN Ia sample described in \cite{Wang2009c}, but consistent with the \cite{Folatelli2013} SN Ia sample and comparable to, for example, SN~2004eo (Fig.~\ref{fig:classificationspectrum}).

Another notable feature in the SEDM spectrum is the complete lack of O\,{\sc i} 7774 \AA\ absorption (Fig.~\ref{fig:classificationspectrum}). Even though O\,{\sc i} absorption in SNe Ia is quite common, in particular 91T-like SNe Ia can have shallow or non-existent O\,{\sc i} \cite{Silverman2012}. This is clearly visible in the matched spectrum of SN 2001V. In contrast, SNe Ibc that lack O\,{\sc i} absorption are extremely uncommon, especially at $\sim$12 days post peak \cite{Liu2016,Fremling2018} as exemplified by Type Ic SN~1994I in Fig.~\ref{fig:classificationspectrum_methods}. In this figure we have also included the Type Ibn SN~2006jc at a phase similar to that of the SEDM spectrum, to highlight that \eyj does not show any sign of He\,{\sc i} emission lines or the quasi-continuum expected for a Type Ibn at this phase.

An absence of oxygen lines is typical for Type Ia-CSM spectra, both as an absorption feature around peak and as emission in later epochs \cite{Silverman2013,fox2015}, as seen in the early and late spectra of PTF11kx in Figs.~\ref{fig:classificationspectrum} and \ref{fig:laterspectra}, respectively. Similarly, the late spectra of \eyj lack any sign of O\,{\sc i} $\lambda$7774 emission (Fig.~\ref{fig:laterspectra}). Other features in the late-time spectra of \eyj that are typical for Type Ia-CSM include prominent broad Ca\,{\sc ii} emission and a high H$\alpha$/H$\beta$ Balmer ratio, which indicates that the emission lines are likely produced through collisional excitation rather than recombination \cite{Silverman2013}. The high S/N spectrum at 251 days shows both H$\alpha$ and H$\beta$ emission, but after correcting for contribution by the host, only H$\alpha$ shows some residual flux related to the transient. We note that at this late phase \eyj is $\sim$4 magnitudes brighter than expected from a normal SN Ia, such as SN~2004eo \cite{Pastorello2007a}, and the spectrum is dominated by CSM-interaction features.

In conclusion, based on its spectral features we classify \eyj as a Type Ia(-CSM) SN. Furthermore, as we discuss in Sect.~\ref{sec:lightcurve}, the light curves of \eyj show strong similarities to those of PTF11kx, the SN that cemented SNe Ia-CSM as a subclass.

\section{Light curve analysis}
\subsection{Light curve fits}\label{sec:lightcurve}

The light curve of \eyj (Fig.~\ref{fig:lightcurve}) can be divided into two phases, similar to its spectral evolution. In the first phase, lasting $\sim$50 days, the light curve follows a fairly typical bell-like shape, peaking at m$\sim$17.2 in both the $r$ band and the ATLAS bands, which at a luminosity distance of 131.4 Mpc (Sect.~\ref{sec:host}) corresponds to M$\sim-$18.4, not accounting for host extinction. During the first phase the light curve has a red $g-r$ color, consistent with the classification spectrum. The second phase, the tail phase from 50 days onward, is characterized by a slowly evolving light curve with spectra that are dominated by CSM interaction. While the $r$-band light curve continues to fade, albeit at a slower rate of $\sim$0.6 mag per 100 days between day 50 and 251, the $g$-band light curve plateaus. This results in a $g-r$ color change to blue (see bottom panel of Fig.~\ref{fig:lightcurve}), which based on the spectra is driven by the pseudo-continuum blueward of 5700~\AA. This Fe\,{\sc ii} feature, typical for CSM-interaction powered spectra, is well traced by the ZTF $g$ band (4100 $-$ 5500 \AA). From 251 days onward, the light curve fades in all bands at a rate of $\sim$1 mag per 100 days.

The transition between the two phases is well captured by the photometry at 50 days, when the decline in $g$ band is abruptly halted and changes to a plateau lasting $\sim$200 days. This divergence of the $g$-band light curve from a smooth decline is likely the epoch where CSM interaction starts contributing (significantly) to the light curve, and where the spectra start to look like those of SNe Ibn. But even though the late spectra may be similar to SNe Ibn, the light curve is unlike those of documented SNe Ibn. SNe Ibn are characterized by uniform rapidly evolving blue light curves \cite{hosseinzadeh2017}, peaking at $M_r\sim-19.5$. There are a handful of long-lived, slowly evolving SNe Ibn reported in the literature, but they are either much brighter than \eyj \cite{pastorello2015V,kool2021} or have a much longer risetime \cite{karamehmetoglu2017}. None of the literature SNe Ibn show a long-duration ($>$ 300 days) slowly evolving light-curve tail like the one observed in \eyj. It is worth noting there have been suggestions in the literature that some SNe Ibn may come from thermonuclear explosions, hidden by a dense CSM \cite{hosseinzadeh2019}. The discovery of \eyj seemingly supports that notion. 

The post-peak decline rates and peak magnitudes of SNe Ia are strongly correlated (the Phillips relation \cite{phillips1993}), with brighter (fainter) SNe Ia evolving slower (faster). We fit the first phase of the multi-band light curves with SNooPy \cite{snoopy}, to determine if the width (stretch) of \eyj is consistent with the expected peak luminosity. The light curve of \eyj up to 50 days is well described by a SN Ia light curve with an adopted stretch of $s_{BV} = 1.2 \pm 0.1$  and an extinction of $E(B-V) =  0.5 \pm 0.1$ mag (adopting a total-to-selective extinction ratio $R_V = 2.0$), resulting in a peak magnitude $\sim$0.06 mag fainter than expected from the Phillips relation. The required line-of-sight extinction is considerable, but is consistent with the host extinction of E(B$-$V) =  $0.54^{+0.14}_{-0.12}$ mag derived from host galaxy Balmer lines (Sect.~\ref{sec:host}). We apply the same fitting method to the light curve of PTF11kx, consisting of published and previously unpublished photometry. For PTF11kx we adopt the same stretch factor of 1.2, and obtain an extinction of E(B$-$V) =  0.27 $\pm$ 0.02 mag, consistent with the extinction A$_V \sim 0.5$ mag derived by \cite{dilday2012}. After correcting for the fitted host extinction, the resulting absolute magnitude light curves of \eyj and PTF11kx are practically identical in $g$ and $r$ band for the first $\sim$45 days, even though the fits are independent (Fig.~\ref{fig:lightcurve_fits}). The $r$-band light curves peak at M$_r \sim -19.3$ for both SNe, consistent with both SNe Ia and SNe Ia-CSM, although both SNe are on the fainter end of the sample of SNe Ia-CSM described by \cite{Silverman2013}. From the light-curve fits we obtain for \eyj rise times in $g$ and $r$ band of 14 $\pm$ 2 and 16 $\pm$ 2 days since discovery, respectively. This is fast for a SN Ia \cite{Miller2020}, but similar to PTF11kx (Fig.~\ref{fig:lightcurve_fits}).

An important caveat about the light curve fit is that the intrinsic decline rate of \eyj could appear slower because of the contribution by CSM interaction. Based on the color evolution of the light curve, we know from day 50 onward that the CSM contribution is significant, but it is reasonable to assume that some CSM interaction already contributes to the light curve at earlier epochs. This means that the stretch parameter we measure should be regarded as an upper limit, and as a result so is the peak luminosity of the fit. SN~2020eyj, but also PTF11kx, are no typical SNe Ia, so the colors and peak magnitude could (to some extent) also be a property intrinsic to the class.

\subsection{Bolometric light curve}\label{sec:bolometric}
The light curve of \eyj around peak has limited photometric coverage, both in wavelength and cadence, which hinders the construction of a precise, full bolometric light curve. Instead, we compute the bolometric light curve based on the SN Ia light curve template fit obtained in Sect.~\ref{sec:lightcurve}, for epochs when the photometry (notably $g$ band) still matches well with the fitted light curve (up to 38 days after first detection, Fig.~\ref{fig:lightcurve_fits}). From the fitted optical light curves, we flux calibrate, correct for host extinction and integrate the spectral time-series from \cite{Hsiao2007} from the UV to the near-IR (1000-25000 \AA).
For the tail phase, we integrate the Keck spectra at 131 and 251 days from 3000 to 10000 \AA,
and apply a linear extrapolation in the UV to zero flux at 2000 \AA. There is little spectroscopic (Fig.~\ref{fig:laterspectra}) and color (Fig.~\ref{fig:lightcurve}) evolution between the Keck spectrum at 251 days and the NOT spectrum at 328 days, so we extend the pseudo-bolometric light curve to the final photometric epoch at 383 days obtained with ALFOSC on the NOT assuming a constant bolometric correction applied to the $g$-band magnitude. Fig.~\ref{fig:model_comparison} shows the bolometric luminosity inferred from the template fit, the Keck spectra, and the final photometric epoch. The template fit to the initial peak included considerable line of sight extinction of E(B$-$V) = 0.5 mag (Sect.~\ref{sec:lightcurve}). To account for the possibility that \eyj may be intrinsically faint rather than a normal SN Ia significantly dust-extincted, we also include a bolometric light curve for E(B$-$V) = 0. Depending on dust extinction, the total integrated energy radiated across the bolometric light curve amounts to $0.6 - 1.2\times10^{50}$ erg.

\section{Host galaxy}\label{sec:host}
The host of \eyj is a faint and compact galaxy with designation SDSS J111147.15+292305.9 (Fig.~\ref{fig:astrometry}). We retrieved science-ready co-added images from the \textit{Galaxy Evolution Explorer} (\textit{GALEX}) general release 6/7 \cite{Martin2005a}, the Sloan Digital Sky Survey data release 9 (SDSS DR 9; \cite{Ahn2012a}), and the Panoramic Survey Telescope and Rapid Response System (Pan-STARRS, PS1) DR1 \cite{Chambers2016} and measured the brightness of the host using \texttt{LAMBDAR} (Lambda Adaptive Multi-Band Deblending Algorithm in R; \cite{Wright2016a}) and the methods described in \cite{Schulze2021a}. We augment this data set with an optical $r$-band image obtained with ALFOSC on the NOT on 2022 May 2 and UV observations from \textit{Swift}/UVOT from 2022 April 27. The photometry on the UVOT images was done with \texttt{uvotsource} in \texttt{HEASoft} and an aperture encircling the entire galaxy (aperture radius $8''$). Table~\ref{tab:hostphot} lists all measurements. We fit the host galaxy SED with the software package Prospector version 0.3 \cite{Johnson2021a} to determine the host galaxy properties.
We assumed a Chabrier initial mass function \cite{Chabrier2003a} and approximated the star formation history (SFH) by a linearly increasing SFH at early times followed by an exponential decline at late times (functional form $t \times \exp\left(-t/\tau\right)$, where $t$ is the age of the SFH episode and $\tau$ is the $e$-folding timescale). The model was attenuated with the \cite{Calzetti2000a} model. The priors were set identical to \cite{Schulze2021a}. The fit resulted in a low host-galaxy mass of $\log (M/M_\odot) = 7.79^{+0.15}_{-0.34}$. 

We obtained a host galaxy spectrum with LRIS/Keck after \eyj had faded from view, at 678 days. We identify unresolved ($\lesssim 150$ km s$^{-1}$) host galaxy lines in the spectrum, such as numerous Balmer lines in emission and absorption, [N\,{\sc ii}] $\lambda\lambda$6548,6583, [O\,{\sc ii}] $\lambda\lambda$3726,3729, [O\,{\sc iii}] $\lambda\lambda$4959,5007 and [S\,{\sc ii}] $\lambda\lambda$6716,6731, based on which we measure a redshift of $z = 0.0297 \pm 0.0001$. Adopting a flat cosmology with $H_0$ = 70 km~s$^{-1}$ Mpc$^{-1}$ and $\Omega_M$ = 0.3, this redshift corresponds to a luminosity distance to \eyj of 131.4 Mpc, which we use throughout this paper. Correcting for MW extinction the adopted distance results in a host galaxy absolute magnitude of M$_r = -15.8$.

Based on the Balmer decrement measured in the host spectrum, we estimate a host extinction with E(B$-$V) =  $0.54^{+0.14}_{-0.12}$ mag, in agreement with the extinction obtained by fitting the light curves of \eyj with a SN Ia template (Sect.~\ref{sec:lightcurve}). The line ratios of log$_{10}$([O\,{\sc iii}] $\lambda 5007$ / H$\beta$) = 0.39 and log$_{10}$([N\,{\sc ii}] $\lambda 6583$ / H$\alpha$) = $-1.26$ put the host galaxy well into the regime of star forming galaxies on the BPT diagram \cite{BPT1981}. Adopting the parameterisation of the empirical oxygen calibration O3N2 by \cite{Marino2013}, we obtain an oxygen abundance of 12 + log(O/H) = 8.14 $\pm$ 0.03. Such a low oxygen abundance is expected for a low mass galaxy \cite{Sanchez2017a}.

The host properties of 16 SNe Ia-CSM were reported in \cite{Prieto2007a, Silverman2013}. These authors concluded that all objects in their samples exploded in star-forming late-type galaxies (spiral and dwarf galaxies) with absolute magnitudes between M$_r=-20.6$ and $-18.1$~mag. The hosts of 3 SNe in this sample evaded detection in archival SDSS images, implying an absolute magnitude of M$_r>-18$~mag. \eyj exploded in a markedly low-luminosity star-forming dwarf galaxy with an absolute $r$-band magnitude of only M$_r = -15.8$~mag (not corrected for host attenuation). However, the modeling of the host galaxy SED and the Balmer decrement reveal non-negligible attenuation of $0<E(B-V)<0.55$~mag ($3 \sigma$ confidence interval from host SED modeling) or $0.2<E(B-V)<1$~mag ($3 \sigma$ confidence interval from the Balmer decrement), which would alleviate the apparent extremeness of the host galaxy.

\section{Dust properties}\label{sec:infrared}

IR emission is commonly observed in interacting SNe, and can be attributed to the condensation of dust in the SN ejecta or in the shocked CSM, or to pre-existing dust in the CSM that is heated radiatively by the SN emission or by the ejecta/CSM shock interaction (e.g., \cite{kotak2004,Mattila2008,fox2011,Fransson2014}). The mid-IR luminosity of \eyj is at a similar level as for the most IR-luminous interacting SNe, such as Type IIn and Ia-CSM SNe, and at 4.5 $\mu$m is 6$-$10 magnitudes brighter than normal Type Ia SNe and $\gtrsim$4 magnitudes brighter than the Type Ibn SN~2006jc (Fig.~\ref{fig:wise}, bottom panel). 

Assuming optically thin dust, the flux $F_{\nu}$ can be written as \cite{Hildebrand1983}:

\begin{equation}
    F_{\nu} = \frac{M_{\textrm{d}}~B_{\nu }(T_{\textrm{d}})~\kappa_{\nu}(a)}{d^2},
\end{equation}

\noindent where $M_{\textrm{d}}$ is the mass of the dust, $B_{\nu }$ the Planck blackbody function, $T_{\textrm{d}}$ the temperature of the dust, $\kappa_{\nu}(a)$ the dust absorption coefficient as function of dust particle radius $a$, and $d$ the distance to the observer. For simplicity, we assume a simple dust population of a single size composed entirely of amorphous carbon with grain size of 0.1 $\mu$m with the corresponding absorption coefficient $\kappa$ as in \cite{Fox2010,Fox2013}, and fit the WISE data to obtain an estimate of the dust temperature and mass. We note that the dust mass depends on assumed grain size, which we can not constrain on the available data. Varying the grain size from 0.01 to 1.0 $\mu$m changes the derived dust mass by an order of magnitude \cite{Fox2010}. Over the first three epochs, up to 412 days, we derive a constant dust temperature of $\sim$800 K (Table~\ref{tab:wise}), consistent with a lack of color evolution in the WISE photometry (Fig.~\ref{fig:lightcurve}). Only at the fourth WISE epoch (614 days) do we see a significant drop in the dust temperature, to $608\pm23$~K. These dust temperatures are well below the expected evaporation temperature of dust (1500~K for silicates and 1900~K for graphite grains, e.g. \cite{Fransson2014}). In addition to the dust temperatures, we obtain dust mass estimates of $(1.8\pm0.3) \times10^{-3}$~\msun~to $(9.9\pm2.1) \times10^{-3}$~\msun~for the first and the final WISE epochs, respectively (Table~\ref{tab:wise}). The dust mass estimated for the final epoch corresponds to a CSM mass of $1\times (0.01/r_{dg})$~\msun, where $r_{dg}$ is the dust-to-gas ratio. The total integrated energy emitted in the IR is $9\times10^{49}$ erg (Table~\ref{tab:wise}), which is similar to the integrated energy emitted in the optical (Sect.~\ref{sec:bolometric}).

In the case of optically thin dust that we consider here, the blackbody radius can be interpreted as a lower limit to the radius at which the dust resides. In the case of \eyj, the blackbody radius is $(2.5\pm0.2) \times10^{16}$~cm in the first epoch, and increases thereafter to $(6.4\pm0.6) \times10^{16}$~cm at 614 days (Table~\ref{tab:wise}). Assuming a SN ejecta velocity of $10^{4}$ km s$^{-1}$, by 59 days the ejecta would only have traveled $\sim20$\% of the distance inferred from the blackbody fit at that epoch. Combined with the constant dust temperature, this suggests that the IR emission of \eyj is dominated by pre-existing dust getting radiatively heated by CSM interaction emission, as was the case in Type Ia-CSM SN 2005gj \cite{Fox2013}. Furthermore, because the dust reached a temperature of 800 K as early as 59 days and showed no significant evolution afterwards, it is unlikely that any surrounding dust was evaporated due to the SN, because such hot dust would have dominated the IR flux. For a peak SN luminosity of $\sim10^{43}$ erg s$^{-1}$ (Fig.~\ref{fig:model_comparison}), the dust evaporation radius is $R_{\textrm{evap}}$ = (0.34--2.6) $\times10^{17}$~cm, depending on dust grain size and composition \cite{Fransson2014}. The lack of dust at the sublimation temperature implies that the immediate region surrounding the SN was devoid of dust, much like the CSM-free cavity inferred from the SN light curve.

The He\,{\sc i} and H$\alpha$ emission line profiles show the red wing being attenuated with time (Fig.~\ref{fig:emlines}). Such an evolution in the line profiles has been interpreted to result from condensation of dust in the ejecta or the shocked CSM, obscuring the red wing of the emission line \cite{Pozzo2004,fox2011,smith2012b}. Similar line profiles have been observed in many SNe Ia-CSM \cite{Silverman2013} and in the prototypical Type Ibn SN 2006jc, where the evolution of the line profiles was attributed to dust condensing in a cool dense shell produced by the interaction of the ejecta with CSM also producing a substantial IR excess \cite{Mattila2008}. Interestingly, such line profile evolution has also been observed in the He nova V445 Pup, where it was attributed to dust obscuration within the shell \cite{Woudt2009}. In particular, for Type Ia-CSM SN~2005gj dust formation was inferred from line profiles \cite{Silverman2013}, while the bulk of the IR emission was also attributed to pre-existing dust \cite{Fox2013}. While the line asymmetry in the spectra \eyj is consistent with dust formation, it must be noted that asymmetric line profiles can also arise from optical depth effects, for example in SNe Ibn \cite{Dessart2022}. A significant contribution to the IR flux by newly formed dust is also not consistent with the lack of color evolution in the light curve of \eyj from day 100 onward. The dust formation in SN~2006jc was accompanied by a reddening of the optical light curve \cite{Mattila2008}, which we do not observe in \eyj past 100 days. So, we attribute the bright IR counterpart of \eyj predominantly to pre-existing dust, similar to the conclusion drawn for the IR counterpart of the prototypical Type Ia-CSM SN~2002ic \cite{kotak2004}.

\section{CSM origin}

\subsection{Optically thick wind}\label{sec:wind}

Using progenitor models for the He star donor SN Ia channel from \cite{wang2009a}, \cite{Moriya2019} investigated the CSM properties resulting from this channel, where accretion from a non-degenerate He star allows the accompanying WD to reach the Chandresekhar limit. The study by \cite{Moriya2019} focused on the low circumstellar density regime, where the CSM properties in the WD + He star systems still adhere to the stringent CSM constraints imposed by radio non-detections of SNe Ia \cite{Horesh2012,pereztorres2014,Chomiuk2016,Lundqvist2020}. Here we explore if the models with sufficiently dense CSM, with a wind-like density profile ($\rho \propto r^{-2}$), can explain the interaction powered light-curve tail of \eyj and the detections at radio wavelengths. In order to quantify the properties of the CSM, we fit the CSM interaction-powered tail of the bolometric light curve using the analytical model from \cite{Moriya2019}, and use the resulting mass-transfer rates to fit the radio detections.

Fig.~\ref{fig:model_comparison} shows the bolometric light curve of \eyj, with the initial peak described by the SN Ia template fit (solid line), and for the tail phase the luminosities inferred from the Keck spectra at 131 and 251 days and the ALFOSC epoch at 383 days (Sect.~\ref{sec:lightcurve}). From the light curve described by the SN Ia component alone (solid and dotted line), it is also clear that the late-time light curve of \eyj cannot be powered by $^{56}$Ni decay, since the flux integrated across the Keck spectrum at 131 days is already at least ten times larger than what the radioactive decay delivers. Also plotted are the CSM-interaction model fits to the light curve tail, both for an E(B-V) = 0 mag and E(B-V) = 0.5 mag as discussed in Sect.~\ref{sec:lightcurve}. Assuming a pre-SN wind velocity of 1000 km s$^{-1}$, the CSM-powered tail of \eyj is consistent with mass-transfer rates between $10^{-3}$~\msun~yr$^{-1}$ (E(B-V) = 0 mag) and $3\times10^{-2}$~\msun~yr$^{-1}$ (E(B-V) = 0.5 mag), which is 1-2 orders of magnitude larger than considered in the original study \cite{Moriya2019}. At these very high mass-transfer rates, the critical mass accretion rate by the WD is exceeded, and the excess is ejected as an optically thick wind resulting in an extended He envelope \cite{Moriya2019}. In the model the forward shock reaches to $\sim10^{17}$ cm in 800 days. If we assume a wind velocity of 1000~km~s$^{-1}$, the CSM mass within $10^{17}$ cm in the models range from 0.3~\msun~to 1~\msun, for E(B-V) = 0 and 0.5 mag, respectively.

Fig.~\ref{fig:model_comparison} shows the wind model radio light curves fitted to the radio detections at 5.1 GHz, adopting $\epsilon_{\textrm{e}}$ = 0.1 and mass-transfer rates of $10^{-3}$~\msun~yr$^{-1}$ and $3\times10^{-2}$~\msun~yr$^{-1}$, for E(B-V) = 0 and 0.5 mag, respectively. We consider both synchrotron emission with synchrotron self-absorption (SSA) and free-free absorption (FFA), but note that at the late phase of the radio detection, FFA only has a minor impact. The radio light curve with an adopted mass-transfer rate of $10^{-3}$~\msun~yr$^{-1}$ is consistent with the radio detections of \eyj at 5.1 GHz, with microphysics parameter $\epsilon_{\textrm{B}} = 1.7\times10^{-3}$. For the high extinction scenario, with a mass-transfer rate of $3\times10^{-2}$~\msun~yr$^{-1}$, the model fits when $\epsilon_{\textrm{B}} = 1.5\times10^{-5}$. In either case, the late time evolution follows the observed power law decline rate of the observed radio luminosity of $\beta = -1.6$, which is comparable to that for hydrogen-free SNe Ibc \cite{Weiler2002}.

It is worth noting that the bolometric light curve only extends to 400 days, whereas the first detection of \eyj at 5 GHz took place at 605 days. Furthermore, it has been argued that the mass-transfer rates associated with the optically thick wind phase ($>10^{-7}$~\msun~yr$^{-1}$) do not lead to SNe Ia, but rather to accretion induced collapse of the WD \cite{Wang2017,Wong2019}, although alternative wind models have been suggested to overcome this problem \cite{Meng2017}.

\subsection{CSM shells}\label{sec:shells}

The CSM surrounding the H-rich analog of \eyj, PTF11kx, was argued to be concentrated in shells \cite{dilday2012}. Other SNe Ia have shown evidence for CSM concentrated in thin shells, albeit at distances ($\sim10^{16}$ cm) that no interaction with the ejecta is expected \cite{Patat2007,Simon2007,Maguire2013,Sternberg2014}. Shells have also been invoked for the configuration of the CSM in core-collapse H-rich Type IIn SNe, and typically attributed to ejection events by their massive progenitors. One noteworthy example is the the well studied SN~2014C, which transitioned from a stripped-envelope SN to a Type IIn SN due to interaction with a distant shell, and was detected in the radio \cite{Milisavljevic2015,Margutti2017}. Models for the radio emission of SNe Ia colliding with a constant-density shell of CSM have been previously presented in the literature, along with approximate functional forms to describe the evolution of the optically thick synchrotron light curve \cite{Harris2021}. Since those models assume hydrogen-rich material, for our calculations we modify $n_e = \rho/m_p$ to $n_e = \rho/(2m_p)$; otherwise we use the default parameters, notably $\epsilon_B=0.1$. 
We explore shell models with a range of CSM masses $M_{\mathrm{csm}}=(0.01-1)~M_\odot$ and interaction end times from $t_\mathrm{end}=328$~days (the spectrum that does not show prominent He~I lines) to $t_\mathrm{end}=763$~days (the second radio detection) -- in this model, interaction must have ended before the second radio detection for the radio emission to have declined between the two observations. 
We assume a shell inner radius of $R_\mathrm{in} = (30,000~\mathrm{km~s^{-1}})(50~\mathrm{days}) = 1.3\times10^{16}~\mathrm{cm}$ to close the system of equations in the model;
then, the ranges of $M_\mathrm{csm}$ and $t_\mathrm{end}$
correspond to a range of shell widths $\Delta R/R_\mathrm{in}=3.4-7.5$.
For each model we calculate the representative model error as $\sigma_\mathrm{mod} = \max(|L_{\nu,\mathrm{obs}}(t_i)-L_{\nu,\mathrm{mod}}(t_i)| / \Delta L_{\nu,\mathrm{obs}}(t_i))$, where subscripts ``obs'' and ``mod'' refer to observed and modeled values, $L_\nu$ is spectral luminosity, and $\Delta L_\nu$ is the error on the luminosity (flux error only; error in distance is not included). The best-fit model by this metric has $M_\mathrm{csm} = 0.36~M_\odot$ and $t_\mathrm{end} = 665~\mathrm{days}$, which is a very similar mass to what is found for PTF11kx based on analysis of its optical spectra \cite{Graham2017}. We find models with $\sigma_\mathrm{mod}\leq3$ have $t_\mathrm{end}\sim(500-763)~\mathrm{days}$ and $M_\mathrm{csm} \sim (0.2-0.5)~M_\odot$, while those with $\sigma_\mathrm{mod}\leq1$ (i.e., a better fit) have $t_\mathrm{end}\gtrsim580~\mathrm{days}$ and $M_\mathrm{csm} \sim (0.3-0.4)~M_\odot$.
The best fit shell model is shown in Fig.~\ref{fig:radio_comparison}.

\subsection{V445 Puppis}\label{sec:v445_pup}

The nova outburst of V445 Pup in the year 2000 lacked any Balmer emission in the spectra of its ejecta, but instead was characterized by He and carbon emission lines \cite{Ashok2003,Kato2003}, making it the first and so far only known He nova system. Based on light curve modeling, a mass ($\geq 1.35$~\msun) close to the Chandresekhar limit was inferred for the WD in V445 Pup \cite{Kato2008}, consistent with the observed high ejecta velocities up to 8450 km s$^{-1}$ \cite{Woudt2009}.  Combined with a high mass-transfer rate $ > 10^{-7}$~\msun~yr$^{-1}$, where half of the accreted matter remains on the WD \cite{Kato2008}, V445 Pup is considered to be a prime candidate progenitor for the single degenerate He + WD SN Ia progenitor channel. 

Based on infrared spectra showing prominent carbon lines \cite{Ashok2003,Iijima2008}, and a rapid decline in the light curve of V445 Pup, it was shown that a carbon-rich thick dust shell must have formed in the nova ejecta \cite{Ashok2003,Kato2003}. High resolution near-IR images resolved the nova event into an expanding narrow bipolar shell with bulk velocities of $\sim$6700 km~s$^{-1}$, and a perpendicular central dust disc that strongly attenuates the optical He\,{\sc i} emission lines arising from the receding shell \cite{Woudt2009}. Seven years after the outburst, the bipolar shell of V445 Pup, as imaged in the near-IR, extended to $\sim10^{17}$ cm, and the central dust torus had an outer radius (perpendicular to the lobes) of $\gtrsim10^{16}$ cm \cite{Woudt2009}. An outer dust shell in a V445 Pup-like system could survive dust sublimation from a SN Ia explosion, depending on peak luminosity and grain composition \cite{Fransson2014}. A recent study of the long-lived radio evolution of V445 Pup showed the system was continuously synchrotron luminous for years after the outburst \cite{Nyamai2021}. The synchrotron emission originated from the inner edge of the equatorial disc, and was interpreted as interaction between a wind coming off the WD from nuclear burning, and the surviving disc. The persistence of the disc through the nova outburst suggests the disc is at least comparable in mass with the mass of the nova ejecta, which was estimated to be $\sim10^{-4}$~\msun \cite{Kato2008}. In turn, the mass of the WD in V445 Pup, close to the Chandrasekhar limit, limits the ejecta mass in the system to not more than $\sim10^{-3}$ \msun\ (\cite{Kato2008}, their Fig. 7).

\subsection{ISM}\label{sec:dd_radio}

Radio emission can potentially arise from a Type Ia SN in the double-degenerate scenario as a result of interaction with the ISM. We have modeled the radio light curve from such a merger scenario in the same way as in \cite{Kundu2017,Lundqvist2020}, i.e., we assume that the supernova is the result of two merging white dwarfs with masses 0.9 and 1.1 \msun\ as described by \cite{Pakmor2012}. The outermost ejecta has a density slope $\propto \rho^{-n}$ with $n$ = 13 (see \cite{Kundu2017} for a discussion on $n$). The microphysics parameters are the standard values $\epsilon_{\textrm{e}}$ = 0.1 and $\epsilon_{\textrm{B}}$ = 0.01. The modeled radio emission increases with time (Fig.~\ref{fig:radio_comparison}), and to agree with the observed 5.1 GHz fluxes at 605 and 741 days, the ISM electron density has to be 660 cm$^{-3}$ and 450 cm$^{-3}$, respectively, assuming fully ionized hydrogen and helium with He/H = 0.1. For $n$ = 13 and fixed $\epsilon_{\textrm{e}}$, the electron density scales roughly as $\epsilon_{\textrm{B}}^{-0.74}$, so other ISM densities are possible accordingly. For a likely upper limit on $\epsilon_{\textrm{B}}$ of 0.1, the ISM density would be $n_{\textrm{e}}$ = 85 cm$^{-3}$ to fit the flux at the second epoch, and for $\epsilon_{\textrm{B}}$ of 0.001, $n_{\textrm{e}}$ = 2570 cm$^{-3}$. The increase in radio flux with time is opposite to what is observed, and is a property for all our ISM models with $n > 7.1$. Lower $n$-values are not expected \cite{Kundu2017}, and the densities required in our ISM models are much higher than normal ISM densities. Moreover, for the $n$ = 13, $\epsilon_{\textrm{B}}$ = 0.01 model, where $n_{\textrm{e}}$ = 450 cm$^{-3}$, the modeled flux for the first epoch undershoots by 2 sigma (Fig.~\ref{fig:radio_comparison}). In summary, our radio observations and their modeling argue strongly against an ISM scenario, which arises from a double degenerate progenitor system. Furthermore, the observed strong helium lines are also at odds with an ISM scenario \cite{Osterbrock2006}. We therefore conclude that \eyj did not result from the thermonuclear runaway of a WD in a DD progenitor system, leaving the SD scenario as the only viable alternative.

\subsection{Precursor search}\label{sec:precursor}
The CSM surrounding \eyj could have originated from one or more novae such as observed in V445 Pup. We investigate if a similar outburst at the location of \eyj\ can be found in ZTF data going back $>2$ years. The position of \eyj\ was observed 772 times (after quality cuts) in the $g$, $r$, and $i$ bands across 202 different nights in the final 2.29 years before the SN explosion. There are no significant pre-explosion detections in unbinned or binned light curves (1-day to 90-day long bins) following the search method described by \cite{strotjohann2021}. When combining observations in week-long bins we reach a median limiting absolute magnitude of $-$14.28 in the $r$ band ($-$14.26 in the $g$ band). We can hence rule out precursors that are brighter than $-$14 magnitude 21\% of the time in the $r$ band (16\% of the time in $g$ band). Precursors brighter than magnitude $-$15 can be ruled out 49\% of the time in $r$ band (39\% for $g$ band) in the final 2.29 years before the SN. The nova outburst of V445 Pup peaked at $m_V = 8.6$ \cite{kato2000}, which at a distance of 8.2 kpc \cite{Woudt2009} equates to an absolute magnitude of $M_V = -1$, far below the detection threshold of ZTF.

\end{methods}
\clearpage

\begin{addendum}

\item The authors thank the anonymous referees for their thoughtful comments and feedback, which have strengthened this manuscript. We thank Luc Dessart, Mike Barlow and Jakob Nordin for helpful feedback and discussions, and Tamas Szalai for providing us with their mid-IR Spitzer light curves.

ECK acknowledges support from the G.R.E.A.T research environment, funded by {\em Vetenskapsr\aa det}, the Swedish Research Council, project number 2016-06012; the research project grant ``Understanding the Dynamic Universe'' funded by the Knut and Alice Wallenberg Foundation under Dnr KAW 2018.0067; and The Wenner-Gren Foundations. 
JM and MPT acknowledge financial support from the State Agency for Research of the Spanish MCIU through the ``Center of Excellence Severo Ochoa'' award to the Instituto de Astrof\'isica de Andaluc\'ia (SEV-2017-0709) and from the grant IAA4SKA (Ref. R18-RT-3082) from the Economic Transformation, Industry, Knowledge and Universities Council of the Regional Government of Andalusia and the European Regional Development Fund from the European Union. JM also acknowledges support from the grant RTI2018-096228-B-C31 (MICIU/FEDER, EU). JM and MPT acknowledge the Spanish Prototype of an SRC (SPSRC) service and support funded by the Spanish Ministry of Science, Innovation and Universities, by the Regional Government of Andalusia and by the European Regional Development Funds. 
TJM is supported by the Grants-in-Aid for Scientific Research of the Japan Society for the Promotion of Science (JP20H00174, JP21K13966, JP21H04997).
LC and CH acknowledge support from NSF grants AST-1751874, AST-1907790, and AST-2107070.
SM acknowledges support from the Academy of Finland project 350458.
AGY’s research is supported by the EU via ERC grant No. 725161, the ISF GW excellence center, an IMOS space infrastructure grant and BSF/Transformative and GIF grants, as well as the André Deloro Institute for Advanced Research in Space and Optics, the Schwartz/Reisman Collaborative Science Program and the Norman E Alexander Family M Foundation ULTRASAT Data Center Fund, The Kimmel center for Planetary Sciences, Minerva and Yeda-Sela;  AGY is the incumbent of the The Arlyn Imberman Professorial Chair. 
NLS is funded by the Deutsche Forschungsgemeinschaft (DFG, German Research Foundation) via the Walter Benjamin program – 461903330.
KM is supported by H2020 European Research Council (ERC) Starting Grant no.~758638 (SUPERSTARS). 
ER has received funding from the European Research Council (ERC) under the European Union’s Horizon 2020 research and innovation programme (grant agreement no 759194 - USNAC)

Based on observations obtained with the Samuel Oschin Telescope 48-inch and the 60-inch Telescope at the Palomar Observatory as part of the Zwicky Transient Facility project. ZTF is supported by the National Science Foundation under Grant No. AST-2034437 and a collaboration including Caltech, IPAC, the Weizmann Institute of Science, the Oskar Klein Center at Stockholm University, the University of Maryland, Deutsches Elektronen-Synchrotron and Humboldt University, the TANGO Consortium of Taiwan, the University of Wisconsin at Milwaukee, Trinity College Dublin, Lawrence Livermore National Laboratories, IN2P3, France, the University of Warwick, the University of Bochum, and Northwestern University. Operations are conducted by COO, IPAC, and UW. 

This work was supported by the GROWTH Marshal project \cite{Kasliwal2019} funded by the National Science Foundation under Grant No 1545949.

SED Machine is based upon work supported by the National Science Foundation under Grant No. 1106171.

The ZTF forced-photometry service was funded under the Heising-Simons Foundation grant \#12540303 (PI: Graham). 

Based on observations made with the Nordic Optical Telescope, owned in collaboration by the University of Turku and Aarhus University, and operated jointly by Aarhus University, the University of Turku and the University of Oslo, representing Denmark, Finland and Norway, the University of Iceland and Stockholm University at the Observatorio del Roque de los Muchachos, La Palma, Spain, of the Instituto de Astrofisica de Canarias.

The data presented here were obtained in part with ALFOSC, which is provided by the Instituto de Astrofisica de Andalucia (IAA) under a joint agreement with the University of Copenhagen and NOT.

The Liverpool Telescope is operated on the island of La Palma by Liverpool John Moores University in the Spanish Observatorio del Roque de los Muchachos of the Instituto de Astrofisica de Canarias with financial support from the UK Science and Technology Facilities Council.

e-MERLIN is a National Facility operated by the University of Manchester at Jodrell Bank Observatory on behalf of STFC.

This work has made use of data from the Asteroid Terrestrial-impact Last Alert System (ATLAS) project. The ATLAS project is primarily funded to search for near earth asteroids through NASA grants NN12AR55G, 80NSSC18K0284, and 80NSSC18K1575; byproducts of the NEO search include images and catalogs from the survey area. This work was partially funded by Kepler/K2 grant J1944/80NSSC19K0112 and HST GO-15889, and STFC grants ST/T000198/1 and ST/S006109/1. The ATLAS science products have been made possible through the contributions of the University of Hawaii Institute for Astronomy, the Queen’s University Belfast, the Space Telescope Science Institute, the South African Astronomical Observatory, and The Millennium Institute of Astrophysics (MAS), Chile.

The Pan-STARRS1 Surveys (PS1) and the PS1 public science archive have been made possible through contributions by the Institute for Astronomy, the University of Hawaii, the Pan-STARRS Project Office, the Max-Planck Society and its participating institutes, the Max Planck Institute for Astronomy, Heidelberg and the Max Planck Institute for Extraterrestrial Physics, Garching, The Johns Hopkins University, Durham University, the University of Edinburgh, the Queen's University Belfast, the Harvard-Smithsonian Center for Astrophysics, the Las Cumbres Observatory Global Telescope Network Incorporated, the National Central University of Taiwan, the Space Telescope Science Institute, the National Aeronautics and Space Administration under Grant No. NNX08AR22G issued through the Planetary Science Division of the NASA Science Mission Directorate, the National Science Foundation Grant No. AST-1238877, the University of Maryland, Eotvos Lorand University (ELTE), the Los Alamos National Laboratory, and the Gordon and Betty Moore Foundation.

We acknowledge ESA Gaia, DPAC and the Photometric Science Alerts Team (\href{http://gsaweb.ast.cam.ac.uk/alerts}{http://gsaweb.ast.cam.ac.uk/alerts}).

\item[Contributions] ECK led the follow-up observations and is the primary author of the manuscript. JJ conducted the spectral analysis and SN Ia light curve modeling, and contributed to the source and infrared analysis. JS contributed significantly to the writing of the manuscript and the source analysis, and conducted follow-up observations with the NOT. JM and MPT led the radio observations and data analysis. TJM, LC, CH and PL conducted the radio light curve modeling. SS conducted the host galaxy analysis. MG conducted follow-up observations with Keck. SM contributed to the writing, and the infrared interpretation. SY contributed to the data analysis. DAP conducted follow-up observations with the Liverpool Telescope. NLS conducted the precursor search. CF, KD and YS conducted follow-up observations. AGY contributed to the writing and source analysis. JL and MS conducted the SEDM spectrum analysis. KM, CO, TMR and SDR contributed to the writing and source analysis. IA, ECB, JSB, SLG, MMK, FJM, MSM, SP, JP, RR, DS are ZTF builders. All authors contributed to edits to the manuscript.

\item[Competing Interests] The authors declare no competing interests.

\item[Correspondence] Correspondence and requests for materials should be addressed to Erik Kool~(email: erik.kool@astro.su.se).

\item[Data Availability] The optical spectra of SN~2020eyj that support the findings of this study have been made available via the WISeREP archive (\url{https://www.wiserep.org/object/14508}). The ZTF photometry is listed in the Supplementary Information. Radio data from the electronic Multi-Element Radio Linked Interferometre Network (e-MERLIN) have been made available on the Zenodo repository with identifier DOI \url{https://doi.org/10.5281/zenodo.7665246}. Data from the NEOWISE-R mission are available from the NASA/IPAC Infrared Science Archive with identifier DOI \url{https://doi.org/10.26131/IRSA124}. Photometry from the Asteroid Terrestrial-impact Last Alert System were obtained from a public source (\url{https://fallingstar-data.com/forcedphot/}).

\item[Code Availability] Upon request, the corresponding author will provide code used to produce the figures. The details of the models used in Sect.~\ref{sec:lightcurve} and Sect.~\ref{sec:wind} can be found in \cite{Kundu2017,Moriya2019,Harris2016,Harris2021} and references therein.

\end{addendum}

\clearpage

\begin{supplement}

\begin{center}
\begin{longtable}{ccccccc}
\caption{Optical photometry of \eyj from ZTF and affiliated programs, in observed magnitudes. Phase is relative to first detection epoch, in rest-frame.}
\label{tab:photometry}\\
\hline\hline
        MJD &   Phase & Filter &     magnitude & error & Limiting magnitude & Telescope+Instrument \\ \hline
58909.278&-5.8&g&-&-&19.93&P48+ZTF\\ 
58909.319&-5.7&r&-&-&20.38&P48+ZTF\\ 
58911.215&-3.9&r&-&-&19.27&P48+ZTF\\ 
58911.402&-3.7&g&-&-&20.25&P48+ZTF\\ 
58912.290&-2.8&g&-&-&19.95&P48+ZTF\\ 
58912.402&-2.7&r&-&-&20.04&P48+ZTF\\ 
58913.235&-1.9&r&-&-&19.59&P48+ZTF\\ 
58913.278&-1.9&g&-&-&19.80&P48+ZTF\\ 
58914.233&-1.0&g&-&-&18.61&P48+ZTF\\ 
58915.212&0.0&g&19.66&0.20&19.29&P48+ZTF\\ 
58915.371&0.2&r&19.11&0.09&19.56&P48+ZTF\\ 
58939.207&23.3&r&17.24&0.01&20.25&P48+ZTF\\ 
58939.209&23.3&r&17.30&0.01&20.28&P48+ZTF\\ 
58939.221&23.3&r&17.23&0.02&19.87&P48+ZTF\\ 
58939.317&23.4&g&17.86&0.03&19.74&P48+ZTF\\ 
58941.180&25.2&r&17.27&0.05&0&P60+SEDM\\ 
58941.250&25.3&g&17.96&0.03&19.86&P48+ZTF\\ 
58954.234&37.9&r&17.65&0.02&20.44&P48+ZTF\\ 
58954.278&37.9&g&18.77&0.04&20.37&P48+ZTF\\ 
58962.245&45.7&g&18.95&0.04&20.63&P48+ZTF\\ 
58962.266&45.7&r&18.11&0.02&20.25&P48+ZTF\\ 
58964.206&47.6&g&19.00&0.04&20.76&P48+ZTF\\ 
58964.207&47.6&g&19.10&0.04&20.57&P48+ZTF\\ 
58965.243&48.6&g&19.03&0.04&20.80&P48+ZTF\\ 
58967.237&50.5&g&19.19&0.05&20.44&P48+ZTF\\ 
58967.238&50.5&g&19.14&0.05&20.37&P48+ZTF\\ 
58968.233&51.5&g&19.04&0.04&20.64&P48+ZTF\\ 
58968.268&51.5&g&19.09&0.04&20.47&P48+ZTF\\ 
58971.265&54.4&g&19.01&0.07&19.60&P48+ZTF\\ 
58973.172&56.3&r&18.39&0.05&19.43&P48+ZTF\\ 
58973.246&56.3&g&19.02&0.09&19.43&P48+ZTF\\ 
58973.247&56.3&g&19.02&0.09&19.54&P48+ZTF\\ 
58976.192&59.2&g&19.02&0.09&19.52&P48+ZTF\\ 
58976.248&59.3&r&18.69&0.07&19.34&P48+ZTF\\ 
58979.229&62.2&g&18.98&0.04&20.57&P48+ZTF\\ 
58985.190&67.9&g&18.95&0.03&20.81&P48+ZTF\\ 
58985.211&68.0&r&18.73&0.03&20.58&P48+ZTF\\ 
58986.206&68.9&g&19.03&0.04&20.63&P48+ZTF\\ 
58991.174&73.7&g&18.95&0.03&20.67&P48+ZTF\\ 
58995.186&77.6&g&18.85&0.03&20.68&P48+ZTF\\ 
58995.187&77.6&g&19.02&0.03&20.61&P48+ZTF\\ 
59001.229&83.5&r&18.87&0.10&19.37&P48+ZTF\\ 
59004.249&86.4&r&19.15&0.11&19.40&P48+ZTF\\ 
59011.240&93.2&r&18.99&0.05&20.14&P48+ZTF\\ 
59014.204&96.1&g&18.88&0.04&20.43&P48+ZTF\\ 
59014.222&96.1&r&19.01&0.05&20.23&P48+ZTF\\ 
59018.228&100.0&r&19.11&0.07&19.78&P48+ZTF\\ 
59021.215&102.9&g&18.98&0.04&20.45&P48+ZTF\\ 
59021.245&102.9&r&19.13&0.06&20.08&P48+ZTF\\ 
59024.207&105.8&g&18.70&0.07&19.70&P48+ZTF\\ 
59024.224&105.8&r&19.03&0.13&19.12&P48+ZTF\\ 
59038.175&119.4&r&18.98&0.08&19.70&P48+ZTF\\ 
59038.179&119.4&r&19.19&0.08&19.88&P48+ZTF\\ 
59038.184&119.4&r&19.17&0.07&19.89&P48+ZTF\\ 
59038.189&119.4&r&19.29&0.09&19.79&P48+ZTF\\ 
59040.184&121.3&r&19.17&0.09&19.64&P48+ZTF\\ 
59040.189&121.3&r&19.09&0.13&19.05&P48+ZTF\\ 
59040.194&121.3&r&19.00&0.23&18.28&P48+ZTF\\ 
59042.175&123.3&r&19.10&0.10&19.44&P48+ZTF\\ 
59042.180&123.3&r&19.13&0.08&19.68&P48+ZTF\\ 
59042.185&123.3&r&19.22&0.09&19.77&P48+ZTF\\ 
59042.190&123.3&r&19.05&0.08&19.67&P48+ZTF\\ 
59044.174&125.2&r&19.07&0.10&19.46&P48+ZTF\\ 
59044.179&125.2&r&19.16&0.09&19.62&P48+ZTF\\ 
59044.184&125.2&r&19.15&0.10&19.60&P48+ZTF\\ 
59044.189&125.2&r&19.17&0.10&19.58&P48+ZTF\\ 
59046.169&127.1&r&19.39&0.19&19.01&P48+ZTF\\ 
59046.174&127.1&r&19.13&0.11&19.43&P48+ZTF\\ 
59046.179&127.2&r&19.21&0.11&19.48&P48+ZTF\\ 
59048.167&129.1&r&18.91&0.10&19.26&P48+ZTF\\ 
59048.172&129.1&r&19.07&0.10&19.52&P48+ZTF\\ 
59048.177&129.1&r&19.22&0.10&19.64&P48+ZTF\\ 
59048.182&129.1&r&19.11&0.08&19.77&P48+ZTF\\ 
59050.166&131.0&r&18.97&0.16&18.66&P48+ZTF\\ 
59050.171&131.0&r&19.02&0.14&19.00&P48+ZTF\\ 
59050.176&131.0&r&19.52&0.18&19.22&P48+ZTF\\ 
59050.181&131.0&r&19.38&0.14&19.31&P48+ZTF\\ 
59129.515&208.1&r&19.62&0.13&19.55&P48+ZTF\\ 
59129.519&208.1&r&19.68&0.16&19.54&P48+ZTF\\ 
59129.524&208.1&r&19.59&0.13&19.54&P48+ZTF\\ 
59129.529&208.1&r&19.69&0.14&19.50&P48+ZTF\\ 
59129.533&208.1&r&19.54&0.15&19.33&P48+ZTF\\ 
59131.517&210.0&r&19.48&0.21&18.91&P48+ZTF\\ 
59131.522&210.0&r&19.64&0.28&18.71&P48+ZTF\\ 
59131.527&210.0&r&19.71&0.29&18.70&P48+ZTF\\ 
59131.531&210.0&r&19.68&0.22&19.24&P48+ZTF\\ 
59131.536&210.0&r&19.34&0.21&18.86&P48+ZTF\\ 
59135.519&213.9&r&19.57&0.11&19.82&P48+ZTF\\ 
59135.524&213.9&r&19.61&0.13&19.75&P48+ZTF\\ 
59135.529&213.9&r&19.43&0.12&19.61&P48+ZTF\\ 
59135.533&213.9&r&19.48&0.13&19.46&P48+ZTF\\ 
59135.538&213.9&r&19.70&0.24&19.03&P48+ZTF\\ 
59137.521&215.8&r&19.68&0.11&19.89&P48+ZTF\\ 
59137.525&215.8&r&19.69&0.13&19.85&P48+ZTF\\ 
59137.530&215.8&r&19.36&0.09&19.74&P48+ZTF\\ 
59137.535&215.8&r&19.38&0.12&19.49&P48+ZTF\\ 
59137.539&215.9&r&19.60&0.20&19.12&P48+ZTF\\ 
59139.520&217.8&r&19.71&0.13&19.80&P48+ZTF\\ 
59139.525&217.8&r&19.47&0.10&19.80&P48+ZTF\\ 
59139.530&217.8&r&19.37&0.10&19.74&P48+ZTF\\ 
59142.497&220.7&g&19.02&0.04&20.36&P48+ZTF\\ 
59142.536&220.7&r&19.48&0.09&20.04&P48+ZTF\\ 
59144.508&222.6&g&19.06&0.04&20.54&P48+ZTF\\ 
59144.538&222.6&r&19.42&0.08&19.99&P48+ZTF\\ 
59150.509&228.4&r&19.66&0.04&-0&P60+SEDM\\ 
59150.512&228.4&g&19.29&0.05&-0&P60+SEDM\\ 
59150.517&228.5&g&19.11&0.05&20.22&P48+ZTF\\ 
59150.546&228.5&r&19.75&0.23&19.19&P48+ZTF\\ 
59152.475&230.4&r&19.27&0.17&19.03&P48+ZTF\\ 
59152.545&230.4&g&19.03&0.08&19.58&P48+ZTF\\ 
59155.487&233.3&r&19.71&0.16&19.48&P48+ZTF\\ 
59157.493&235.2&r&19.62&0.14&19.48&P48+ZTF\\ 
59157.510&235.2&g&19.17&0.13&19.28&P48+ZTF\\ 
59159.492&237.2&r&19.75&0.12&19.71&P48+ZTF\\ 
59159.524&237.2&g&19.02&0.06&19.88&P48+ZTF\\ 
59165.433&242.9&r&19.66&0.10&19.98&P48+ZTF\\ 
59165.517&243.0&g&19.04&0.04&20.49&P48+ZTF\\ 
59168.509&245.9&g&19.12&0.05&20.38&P48+ZTF\\ 
59168.529&245.9&r&19.86&0.09&20.33&P48+ZTF\\ 
59169.195&246.6&i&20.08&0.22&20.95&LT+IOO\\ 
59169.196&246.6&r&19.67&0.12&20.94&LT+IOO\\ 
59169.199&246.6&z&18.61&0.16&20.96&LT+IOO\\ 
59170.487&247.8&g&19.11&0.04&20.55&P48+ZTF\\ 
59170.547&247.9&r&19.84&0.10&20.24&P48+ZTF\\ 
59173.486&250.8&g&19.17&0.06&20.32&P48+ZTF\\ 
59175.516&252.7&r&19.41&0.26&18.80&P48+ZTF\\ 
59175.529&252.7&g&18.97&0.16&18.90&P48+ZTF\\ 
59176.477&253.7&i&19.32&0.11&19.59&P48+ZTF\\ 
59177.531&254.7&g&19.21&0.05&20.50&P48+ZTF\\ 
59180.484&257.5&g&19.38&0.15&19.17&P48+ZTF\\ 
59180.546&257.6&r&19.84&0.18&19.46&P48+ZTF\\ 
59181.411&258.4&r&19.89&0.21&19.37&P48+ZTF\\ 
59181.442&258.5&i&19.57&0.23&19.01&P48+ZTF\\ 
59182.429&259.4&r&19.64&0.13&19.77&P48+ZTF\\ 
59182.488&259.5&r&19.96&0.21&19.51&P48+ZTF\\ 
59182.562&259.6&g&19.12&0.09&19.63&P48+ZTF\\ 
59183.384&260.4&g&19.62&0.32&18.34&P48+ZTF\\ 
59183.527&260.5&g&19.46&0.05&20.43&P48+ZTF\\ 
59184.379&261.3&g&19.46&0.08&19.90&P48+ZTF\\ 
59184.445&261.4&g&19.04&0.10&19.35&P48+ZTF\\ 
59184.508&261.5&r&19.88&0.16&19.69&P48+ZTF\\ 
59185.507&262.4&g&19.11&0.09&19.64&P48+ZTF\\ 
59185.508&262.4&g&19.17&0.08&19.75&P48+ZTF\\ 
59185.527&262.4&r&19.88&0.15&19.84&P48+ZTF\\ 
59185.528&262.4&r&19.89&0.13&19.83&P48+ZTF\\ 
59186.542&263.4&g&19.58&0.24&18.87&P48+ZTF\\ 
59187.464&264.3&r&19.87&0.25&19.20&P48+ZTF\\ 
59187.533&264.4&g&19.19&0.11&19.42&P48+ZTF\\ 
59187.533&264.4&g&19.30&0.11&19.44&P48+ZTF\\ 
59192.506&269.2&r&19.76&0.21&19.29&P48+ZTF\\ 
59192.523&269.2&r&20.16&0.09&21.57&LCOGT1m+Sinistro\\ 
59192.523&269.2&r&20.12&0.09&21.57&LCOGT1m+Sinistro\\ 
59192.525&269.2&i&20.17&0.13&21.26&LCOGT1m+Sinistro\\ 
59194.499&271.2&r&19.92&0.08&20.48&P48+ZTF\\ 
59194.532&271.2&g&19.37&0.05&20.52&P48+ZTF\\ 
59198.475&275.0&r&19.86&0.10&20.10&P48+ZTF\\ 
59198.506&275.0&g&19.31&0.05&20.58&P48+ZTF\\ 
59199.488&276.0&i&19.31&0.11&19.48&P48+ZTF\\ 
59200.466&276.9&g&19.28&0.13&19.45&P48+ZTF\\ 
59200.517&277.0&r&19.85&0.22&19.36&P48+ZTF\\ 
59202.470&278.9&i&19.51&0.12&19.57&P48+ZTF\\ 
59203.484&279.9&r&20.16&0.15&20.08&P48+ZTF\\ 
59205.431&281.8&g&19.36&0.06&20.46&P48+ZTF\\ 
59215.422&291.5&g&19.32&0.24&18.74&P48+ZTF\\ 
59216.500&292.5&i&20.28&0.39&19.09&P48+ZTF\\ 
59219.394&295.3&i&19.65&0.17&19.38&P48+ZTF\\ 
59219.421&295.3&g&19.51&0.12&19.69&P48+ZTF\\ 
59219.476&295.4&r&20.17&0.15&20.07&P48+ZTF\\ 
59221.404&297.3&r&20.31&0.17&20.10&P48+ZTF\\ 
59221.478&297.3&g&19.54&0.08&20.18&P48+ZTF\\ 
59222.435&298.3&i&19.56&0.15&19.52&P48+ZTF\\ 
59223.401&299.2&g&19.58&0.09&20.10&P48+ZTF\\ 
59223.463&299.3&r&20.45&0.22&19.87&P48+ZTF\\ 
59225.412&301.2&r&20.15&0.16&20.04&P48+ZTF\\ 
59225.435&301.2&i&19.76&0.20&19.35&P48+ZTF\\ 
59225.443&301.2&g&19.67&0.09&20.20&P48+ZTF\\ 
59228.436&304.1&g&19.57&0.06&20.45&P48+ZTF\\ 
59228.476&304.1&r&20.24&0.10&20.54&P48+ZTF\\ 
59228.532&304.2&i&20.07&0.18&19.70&P48+ZTF\\ 
59230.443&306.0&r&20.32&0.14&20.37&P48+ZTF\\ 
59230.464&306.1&g&19.53&0.06&20.66&P48+ZTF\\ 
59231.347&306.9&i&19.86&0.17&19.56&P48+ZTF\\ 
59232.435&308.0&g&19.45&0.05&20.90&P48+ZTF\\ 
59232.485&308.0&r&20.34&0.13&20.34&P48+ZTF\\ 
59248.344&323.4&i&20.05&0.21&19.59&P48+ZTF\\ 
59249.423&324.5&g&19.84&0.16&19.75&P48+ZTF\\ 
59250.454&325.5&g&19.84&0.09&20.20&P48+ZTF\\ 
59251.349&326.3&i&19.91&0.18&19.53&P48+ZTF\\ 
59251.402&326.4&g&19.77&0.07&20.73&P48+ZTF\\ 
59251.446&326.4&r&20.76&0.17&20.50&P48+ZTF\\ 
59253.351&328.3&r&20.51&0.14&20.41&P48+ZTF\\ 
59253.361&328.3&g&19.80&0.08&20.62&P48+ZTF\\ 
59254.361&329.3&i&20.02&0.14&19.93&P48+ZTF\\ 
59255.363&330.2&g&19.90&0.07&20.79&P48+ZTF\\ 
59255.402&330.3&r&20.63&0.12&20.74&P48+ZTF\\ 
59256.390&331.2&r&20.82&0.13&20.80&P48+ZTF\\ 
59257.376&332.2&g&19.87&0.13&20.00&P48+ZTF\\ 
59257.415&332.2&r&20.80&0.26&20.12&P48+ZTF\\ 
59258.327&333.1&r&20.38&0.18&20.01&P48+ZTF\\ 
59260.338&335.1&i&20.63&0.32&19.66&P48+ZTF\\ 
59262.347&337.0&r&20.65&0.15&20.55&P48+ZTF\\ 
59262.409&337.1&g&19.81&0.07&20.59&P48+ZTF\\ 
59264.306&338.9&r&20.47&0.18&20.12&P48+ZTF\\ 
59264.354&339.0&g&19.75&0.06&20.80&P48+ZTF\\ 
59264.383&339.0&i&20.35&0.17&19.98&P48+ZTF\\ 
59266.332&340.9&r&20.89&0.39&19.62&P48+ZTF\\ 
59266.415&341.0&g&19.95&0.13&20.00&P48+ZTF\\ 
59267.270&341.8&r&20.71&0.31&19.87&P48+ZTF\\ 
59267.286&341.8&g&19.99&0.18&19.69&P48+ZTF\\ 
59267.309&341.8&r&20.73&0.31&19.74&P48+ZTF\\ 
59267.351&341.9&i&20.48&0.37&19.38&P48+ZTF\\ 
59267.360&341.9&g&19.79&0.15&19.69&P48+ZTF\\ 
59268.171&342.7&g&20.00&0.25&19.39&P48+ZTF\\ 
59268.276&342.8&r&20.67&0.23&19.95&P48+ZTF\\ 
59268.353&342.9&r&20.71&0.26&20.02&P48+ZTF\\ 
59268.357&342.9&r&21.17&0.39&20.08&P48+ZTF\\ 
59268.430&342.9&g&19.88&0.13&20.06&P48+ZTF\\ 
59268.490&343.0&g&19.78&0.10&20.17&P48+ZTF\\ 
59269.234&343.7&g&20.20&0.27&19.60&P48+ZTF\\ 
59269.310&343.8&r&20.91&0.33&19.81&P48+ZTF\\ 
59270.382&344.8&g&19.86&0.23&19.36&P48+ZTF\\ 
59273.309&347.7&g&19.76&0.36&18.64&P48+ZTF\\ 
59273.379&347.7&g&19.75&0.38&18.64&P48+ZTF\\ 
59274.479&348.8&g&19.37&0.21&18.90&P48+ZTF\\ 
59275.213&349.5&g&19.92&0.21&19.55&P48+ZTF\\ 
59275.339&349.6&r&20.92&0.39&19.76&P48+ZTF\\ 
59278.265&352.5&g&20.10&0.11&20.37&P48+ZTF\\ 
59278.361&352.6&i&20.25&0.28&19.48&P48+ZTF\\ 
59278.412&352.6&r&21.13&0.40&19.95&P48+ZTF\\ 
59281.254&355.4&r&20.80&0.18&20.55&P48+ZTF\\ 
59291.239&365.1&r&21.15&0.25&20.45&P48+ZTF\\ 
59291.269&365.1&i&21.07&0.40&19.83&P48+ZTF\\ 
59291.286&365.1&g&20.31&0.13&20.38&P48+ZTF\\ 
59293.236&367.0&r&21.27&0.34&20.33&P48+ZTF\\ 
59293.298&367.1&g&20.22&0.11&20.63&P48+ZTF\\ 
59294.296&368.0&i&20.48&0.23&19.72&P48+ZTF\\ 
59308.920&382.2&r&21.16&0.11&21.24&NOT+ALFOSC\\ 
59308.930&382.3&i&21.34&0.09&21.25&NOT+ALFOSC\\ 
59308.940&382.3&g&20.21&0.12&21.61&NOT+ALFOSC\\ 
\hline
\end{longtable}
\end{center}

\end{supplement}

\end{document}